\documentclass[twocolumn,
showpacs,
preprintnumbers,
amsmath,amssymb,
aps,
prb,
citeautoscript,
superscriptaddress]{revtex4}
\usepackage[latin1]{inputenc}
\usepackage{graphicx}
\def\tlse{Laboratoire de Physique Th\'eorique, Universit\'e Paul Sabatier, CNRS,
118 Route de Narbonne, 31400 Toulouse, France.}
\def\lyon{Laboratoire de Physique de l'Ecole Normale
Sup\'erieure de Lyon, CNRS, 46 All\'ee d'Italie, 69007 Lyon, France.}
\def\fig{{\small FIG.}}
\newcommand{\bc}{\begin{center}}
\newcommand{\ec}{\end{center}}
\def\Ham{\mathcal{H}}
\def\nh{n_h}
\def\P{\mathcal{P}}
\def\S{\mathbf{S}}
\def\H{\textrm{H}}

\newcommand{\moy}[1]{\left\langle{#1}\right\rangle}
\def\ups{\uparrow}
\def\downs{\downarrow}

\begin{document}

\title{Bosonization and density-matrix renormalization group studies of
the Fulde-Ferrell-Larkin-Ovchinnikov phase and irrational magnetization
plateaus in coupled chains}

\author{G.\ Roux}
\email{roux@irsamc.ups-tlse.fr}
\affiliation{\tlse}
\author{E.\ Orignac}
\email{Edmond.Orignac@ens-lyon.fr}
\affiliation{\lyon}
\author{P.\ Pujol}
\affiliation{\tlse}
\author{D.\ Poilblanc}
\affiliation{\tlse}

\date{\today}

\pacs{71.10.Pm, 74.20.Mn, 74.81.-g, 75.40.Mg}

\begin{abstract}
We review the properties of two coupled fermionic chains, or ladders,
under a magnetic field parallel to the lattice plane. Results are
computed by complementary analytical (bosonization) and numerical
(density-matrix renormalization group) methods which allows a
systematic comparison. Limiting cases such as coupled bands and
coupled chains regimes are discussed. We particularly focus on the
evolution of the superconducting correlations under increasing field
and on the presence of irrational magnetization plateaus. We found the
existence of large doping-dependent magnetization plateaus in the
weakly-interacting and strong-coupling limits and in the non-trivial
case of isotropic couplings. We report on the existence of extended
Fulde-Ferrell-Larkin-Ovchinnikov phases within the isotropic t-J and
Hubbard models, deduced from the evolution of different observables
under magnetic field. Emphasis is put on the variety of
superconducting order parameters present at high magnetic field. We
have also computed the evolution of the Luttinger exponent
corresponding to the ungapped spin mode appearing at finite
magnetization. In the coupled chain regime, the possibility of having
polarized triplet pairing under high field is predicted by
bosonization.
\end{abstract}
\maketitle

\section{Introduction}

Low-dimensional strongly correlated systems have attracted strong
interest in the last years because fluctuations and energetic
competitions drive these systems into exotic
phases. Quasi-one-dimensional or strongly anisotropic two-dimensional
superconductors are also known to be good candidates for the
realization of Fulde-Ferrell-Larkin-Ovchinnikov (FFLO)
phases\cite{FFLO, Ishiguro2002, Dupuis1995, Shimahara1997} because the
orbital effects of the magnetic field, which induce the $\H_{c2}$
critical field in type-II superconductors, can be strongly
suppressed. In a singlet superconductor without orbital effect, there
is a competition between polarizing the spins of the electrons and
binding them into Cooper pairs leading to a theoretical critical field
called the Pauli limit or the paramagnetic limit\cite{Ishiguro2002,
Dupuis1995, Shimahara1997, Sigrist2005}. However, this limit can be
exceeded with an inhomogeneous order parameter (FFLO) which can be
energetically favorable, allowing pockets of polarized electrons and
paired electrons. Qualitatively, the FFLO mechanism consists in giving
singlet Cooper pairs a finite momentum which leads to this
inhomogeneous superconducting order parameter. Among these intriguing
low-dimensional systems, ladders, which consist of a few coupled
chains, proved to display deep new physical behaviors and sustained
considerable experimental and theoretical work\cite{LadderReview}.

At half-filling, ladders are Mott insulators and have a spin gap if
the number of chains is even (this gap goes to zero in the limit of an
infinite number of coupled chains) and no spin gap if it is odd. The
two-leg ladder has thus a spin-liquid ground state with exponentially
decaying magnetic correlations. Spin gaps can also open under magnetic
field for rational values of the magnetization per site leading to
plateaus in the magnetization curve\cite{Cabra1997}. Experimental
evidences of zero magnetization plateaus have been reported on ladder
and coupled-dimer compounds\cite{Chaboussant, Watson2001, Landee2001}
such as Cu$_2$(C$_5$H$_{12}$N$_2$)$_2$Cl$_4$,
(C$_5$H$_{12}$N)$_2$CuBr$_4$ and (5IAP)$_2$CuBr$_4$$\cdot$2H$_2$O.
Away from half-filling, a few systems are known to develop irrational
magnetization plateaus controlled by hole
concentration\cite{Frahm1999, Cabra2000, Cabra2001}. Furthermore, when
holes are introduced into the spin-liquid two-leg ladder, these charge
carriers generically bring the system into either a metallic or a
superconducting state. The isotropic two-leg ladder is known to have a
wide superconducting phase\cite{Hayward1995, Poilblanc2003} and also a
metallic phase with dominant charge-density wave\cite{White2002}
(CDW) fluctuations. Another interesting feature is the appearance of commensurate
CDW for a commensurate hole concentration\cite{White2002}. The
theoretically proposed framework to account for superconductivity in
doped ladders relies on magnetic fluctuations and is based on the
resonating-valence-bond (RVB) mechanism for superconductivity proposed
by Anderson in the context of high-$T_c$
superconductors\cite{Anderson1987}.  Within the isotropic Hubbard and
t-J models, singlet pairing with an unconventional modified $d$-wave
structure is found. The competition\cite{Gozar2005} between
superconductivity and CDW has indeed been observed in the copper oxide
ladder compound Sr$_{14-x}$Ca$_x$Cu$_{24}$O$_{41 + \delta}$ (SCCO) for
which the superconducting state only appears under high
pressure\cite{Uehara1996}. However, the mechanism responsible for
superconductivity in SCCO has not reached a full agreement yet. In
this context, the upper critical magnetic fields determined from
transport measurements\cite{Braithwaite2000, Nakanishi2005} suggest
that the Pauli limit is exceeded in SCCO which reassesses the issue of
the nature of the pairing. Note that superconductivity has been
discovered in the zig-zag ladder subsystem of the copper oxide compound
Pr$_2$Ba$_4$Cu$_7$O$_{15-\delta}$ at ambient
pressure\cite{Matsukawa2004}.

Recently, a superconducting two-leg t-J ladder under a strong magnetic
field in the plane of the ladder was studied
numerically\cite{Roux2006} using the density-matrix renormalization
group\cite{White1992, Schollwock2005} (DMRG) method. The magnetic
curve displays a doping-dependent magnetization plateau, as
predicted\cite{Cabra2002} by Cabra {\it el al.}  for a Hubbard
ladder. In addition to this non trivial magnetic behavior, an
exceeding of the Pauli limit was found. Within the t-J model, this
exceeding of the Pauli limit was explained by a one-dimensional
analogue of the FFLO phase, hence reconciling the expectation of
singlet pairing and the exceeding of the Pauli limit. Lastly, the
behavior of the superconducting correlations studied in
Ref.~[\onlinecite{Roux2006}] showed a surprising behavior in and
outside the plateau with a notable emergence of $S^z=0$ triplet
superconducting correlations.

In this paper, we propose to greatly complete Ref.~[\onlinecite{Roux2006}]
by an extensive comparison of bosonization and numerical
calculations and by extending the results to discuss the case of
weakly-interacting or strongly anisotropic ladders and also to the
case of weakly-coupled chains. In particular, we show that the
picture developed in Ref.~[\onlinecite{Roux2006}] is consistent
with bosonization and extends to both Hubbard and t-J models, supporting its
generality. While most studies on the FFLO phase resort to mean-field
theories for low-dimensional and unconventional
superconductors\cite{Shimahara1997}, we here use approaches which are
more relevant for quasi one-dimensional strongly-correlated systems. In
the same spirit, an early study on the superconducting phase of the
t-J chain which gave evidence of developing FFLO-like
correlations\cite{Pruschke1992} was based on exact diagonalization
computations, and bosonization has also been used in this
context\cite{Yang2001} on an attractive Hubbard system.

The article is organized as follow: first, after introducing
microscopic models (Sec.~\ref{sec:models}), we briefly examine the
non interacting and strong-coupling limits in
Sec.~\ref{sec:non-interacting} where large doping-dependent
magnetization plateaus can occur. The coupled bands regime is
discussed in detail in Sec.~\ref{sec:band-models} and corresponds to
the case of doped isotropic ladders which is the most studied at zero
magnetic field. Lastly, the coupled chains regime is studied under
magnetic field (Sec.~\ref{sec:coupled-chain}).

\section{Microscopic models and convention}
\label{sec:models}

We describe the ladder system with a standard one-orbital Hubbard
model which can have different hopping terms along the chains
($\parallel$) and between the chains ($\perp$). We consider a
situation where no magnetic orbital effect is present and thus only
keep a Zeeman coupling to the spin degree of freedom. This would
experimentally correspond to the situation where the magnetic field
$\H$ is in the plane of the ladder (and even along the direction of
the ladder to minimize all orbital effects). Then, one can write the
Hubbard Hamiltonian
\begin{eqnarray}
\Ham &=& -t_{\parallel}\sum_{i,p=1,2,\sigma} [c^{\dag}_{i+1,p,\sigma}
c_{i,p,\sigma} + h.c. ]\nonumber \\
&& - t_{\perp}\sum_{i,\sigma} [c^{\dag}_{i,2,\sigma} c_{i,1,\sigma} +
 h.c. ] \nonumber \\
&& + U\sum_{i,p=1,2} n_{i,p,\ups} n_{i,p,\downs}  \nonumber \\
&& - \sum_{i,p=1,2} \H \cdot \S_{i,p}
\label{eq:HubbardHam}
\end{eqnarray}
where $c^{\dag}_{i,p,\sigma}$, $\S_{i,p}$ and $n_{i,p,\sigma}$ are
respectively electron creation, spin and density operators at site $i$
on chain $p$ and $\sigma$ is the spin index. The $g\mu_B$ prefactor
has been absorbed in the definition of $\H$ for convenience. Free
bands dispersion is given by
\begin{figure}[t]
\centering
\includegraphics[width=5cm,angle=270,clip]{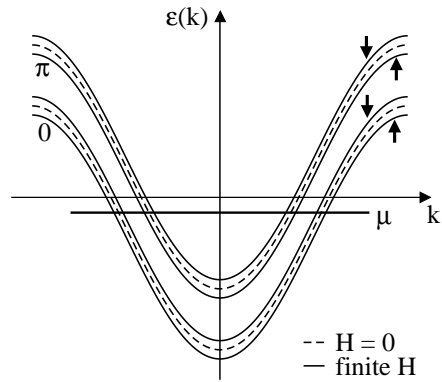}
\caption{Splitting of the bands dispersion in a non-interacting doped
ladder ($\mu$ is the chemical potential) due to Zeeman effect at low
magnetic field $\H$.}
\label{fig:schema}
\end{figure}
\begin{equation}
\label{eq:free-bands}
\varepsilon_{k_y,\sigma}(k) = -2 t_{\parallel} \cos(k) - t_\perp
\cos(k_y) - \H \frac \sigma 2
\end{equation}
(with $k=k_x$ for simplicity and $k_y=0,\pi$) and are sketched in 
\fig~\ref{fig:schema}. If we denote by $k_{F,k_y}^{\sigma}$ the Fermi
wave-vectors, we have the Luttinger sum rule
\begin{equation}
\label{eq:Luttinger}
n = \frac{1}{2\pi} \sum_{(k_y,\sigma)\,\textrm{occ.}} k_{F,k_y}^{\sigma}\;,
\end{equation}
where $n = N^e / (2L) = 1-\delta$ is the electron density and $\delta$
the hole density and $L$ the length of the ladder. Experimentally,
copper-oxide systems have a fixed hole doping $\delta$ rather than a
fixed chemical potential $\mu$ so $\delta$ will be kept fixed in this
paper. Similarly, the magnetization per site $m = ( N^{\ups} -
N^{\downs} ) / (2L)$ satisfies
\begin{equation}
m = \frac{1}{2\pi} \sum_{(k_y,\sigma)\,\textrm{occ.}} \sigma k_{F,k_y}^{\sigma}.
\label{eq:m}
\end{equation}
Relations (\ref{eq:Luttinger}) and (\ref{eq:m}) are also valid in the
presence of interactions. For quasi one-dimensional interacting
systems, a useful theorem is the generalization of the
Lieb-Schultz-Mattis theorem to doped and magnetized states done by
Yamanaka-Oshikawa-Affleck\cite{Yamanaka1997} (YOA). This theorem
relates gap openings to commensurability conditions. When dealing with
spinful fermions, the demonstration is based on the definition of 2
twist operators for spins $\ups,\downs$ which gives two
commensurability conditions for parameters in each sector.  In the
case of doped two-leg ladders, YOA parameters take the simple
form\cite{Cabra2002} $1 -\delta +\sigma m$. If a parameter is
non-integer, the spectrum has gapless excitations in this sector. If a
parameter is non-integer but rational, a gap can open in the sector
(\emph{depending} on the relevance of interactions) but necessarily
the ground-state is degenerate with translational symmetry
breaking. When the parameter is an integer, the sector is gapless or
it is gapped but with a non-degenerate ground state. It is
interesting to note that for $m, \delta \neq 0$, the sector $\ups$ can
get gapped while the sector $\downs$ remains gapless (which
corresponds to the $m = \delta$ plateau phase in
Sec.~\ref{sec:band-models}) so that the usual spin mode-charge mode
separation present when $m=0$ is no longer valid.

In the strong-coupling limit of the Hubbard model $U \gg t$ and for
small hole doping, Hamiltonian (\ref{eq:HubbardHam}) reduces to the
t-J model:
\begin{eqnarray}
\Ham &=& -t_{\parallel}\sum_{i,p=1,2,\sigma}  \P[c^{\dag}_{i+1,p,\sigma}
c_{i,p,\sigma} + h.c. ] \P \nonumber \\
&& -t_{\perp}\sum_{i,\sigma}  \P[c^{\dag}_{i,2,\sigma} c_{i,1,\sigma} +
 h.c. ] \P \nonumber \\
&& +J_{\parallel} \sum_{i,p=1,2} [ \S_{i,p}\cdot\S_{i+1,p} - 
\frac{1}{4} n_{i,p} n_{i+1,p}] \\
&& +J_{\perp} \sum_{i} [ \S_{i,1}\cdot\S_{i,2} - 
\frac{1}{4} n_{i,1} n_{i,2}] \nonumber 
 - \sum_{i,p=1,2} \H \cdot \S_{i,p}\;,
\label{eq:tJHam}
\end{eqnarray}
in which $\P$ is the Gutzwiller projector, $n_{i,p} = \sum_{\sigma}
n_{i,p,\sigma}$ and the same convention is used to label the
antiferromagnetic couplings $J_{\parallel, \perp} = 4 t_{\parallel,
\perp}^2 / U$.  In what follows, these two microscopic models will be
studied numerically and the isotropic model will assume $t =
t_{\parallel} = t_{\perp}$ and $J = J_{\parallel} = J_{\perp}$.

Bosonization allows to study the low-energy properties and correlation
functions of 1D-like systems using field theoretical methods. The
bosonization Hamiltonians used to describe ladders\cite{nagaosa_2ch,
khveshenko_2chain, finkelstein_2ch, schulz_2chains, balents_2ch} fall
into two classes, coupled bands models to be reviewed in
Sec.~\ref{sec:band-models} and coupled chains models to be reviewed in
Sec.~\ref{sec:coupled-chain}. The first approach is more appropriate
to study the system with isotropic parameters. There is however the
possibility, by strongly reducing the interchain hopping amplitude
$t_{\perp}$, to obtain a cross-over to a regime where the coupled
chain model is best suited to describe the behavior of the system
(Sec.~\ref{sec:coupled-chain}).

\section{Non interacting and strong-coupling limits}
\label{sec:non-interacting}
\begin{figure}[t]
\centering
\includegraphics[width=3.4cm,angle=270,clip]{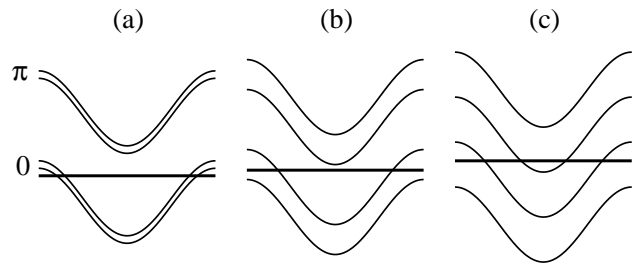}   
\caption{Bands evolution corresponding to
\fig~\ref{fig:strong-coupling-free}.}
\label{fig:free-evolution}
\end{figure}

\begin{figure}[t]
\centering
\includegraphics[width=5cm,angle=270,clip]{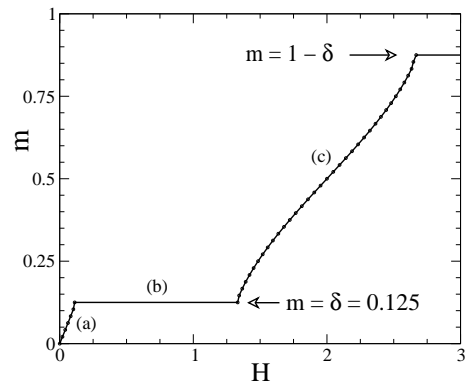}   
\caption{Magnetization curves for the free system with
$t_{\parallel}=0.2$ and $t_{\perp}=1.0$ showing a plateau
at $m=\delta$ of width $2(t_{\perp}-2t_{\parallel})$.}
\label{fig:strong-coupling-free}
\end{figure}

\emph{Non interacting system.}  It is interesting to discuss first the
non-interacting system using the relations (\ref{eq:free-bands}) and
keeping $\delta$ fixed instead of the chemical potential. Two main
cases are possible: either a strong interband coupling with $t_{\perp}
> 2t_{\parallel}$ or a small interband coupling for $t_{\perp} <
2t_{\parallel}$.

In the first case, the only bands which are partially filled at low
magnetic field are the up and down spins bonding bands $(0,\ups)$ and
$(0,\downs)$ (see \fig~\ref{fig:free-evolution} {\bf(a)}). A first
critical field corresponds to the complete filling of $(0,\ups)$
(\fig~\ref{fig:free-evolution} {\bf(b)}). This induces a plateau with
$m=\delta$ in the magnetization curve. This plateau is
doping-dependent and similar to what has been predicted for other
doped systems\cite{Cabra2000,Cabra2001}. The width of the plateau can
be deduced from energetic considerations\cite{Cabra2001} to be
$2(t_{\perp}-2t_{\parallel})$ in our case. When the plateau ends
(\fig~\ref{fig:free-evolution} {\bf(c)}), the band $(\pi,\ups)$ starts
to be partially filled.  When the band $(0,\downs)$ gets empty, all
electrons are polarized and the magnetization curve is constant with
value $m=1-\delta$. For open boundary conditions (OBC), a $m=\delta$
plateau is also found as displayed in
\fig~\ref{fig:strong-coupling-free} (see Sec.~\ref{sec:plateaus} for
computational details). The width is in good agreement with
$2(t_{\perp}-2t_{\parallel})$. Note that this plateau does not
originate from interactions contrary to what will be discussed in the
strong-coupling approach below and in Sec.~\ref{sec:plateaus}, but is
simply due to band-filling effects.

In the second case, all four bands are partially filled at low
magnetic field. No plateaus are found but filling or emptying
successively the bands will induce cusps in the magnetization curve
and a decrease of the slope of the magnetization because fewer
electrons contribute to the magnetization. First, the band
$(\pi,\downs)$ gets empty, next $(0,\ups)$ becomes completely filled
and lastly, $(0,\downs)$ gets empty corresponding to the saturation of
the magnetization. An example of such a curve for $t_{\parallel} =
t_{\perp}$ with OBC is given on \fig~\ref{fig:magnetization}. Two
cusps are visible in this curve as well as a decrease of the average
slope of the magnetization.

\begin{figure}[t]
\centering
\includegraphics[width=5cm,angle=270,clip]{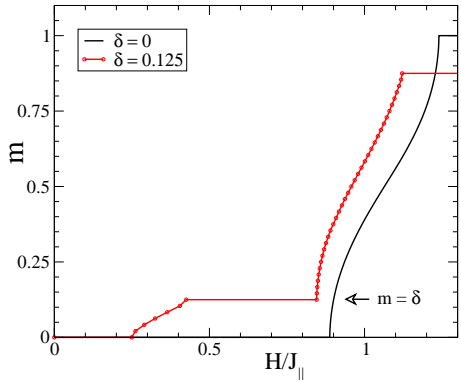}
\caption{(Color online) Plateaus in the strong-coupling limit. We
chose $J_{\perp} = 2.5$, $J_{\parallel} = 0.3$ and
$t_{\parallel}=t_{\perp}=1.0$. The $m=\delta$ plateau is an effect of
interactions and the ground-state in this plateau has unpaired
holes. Still, holes are paired for magnetizations with $m\leq
m_c<\delta$.}
\label{fig:strong-coupling}
\end{figure}

\begin{figure}[t]
\centering
\includegraphics[width=5cm,angle=270,clip]{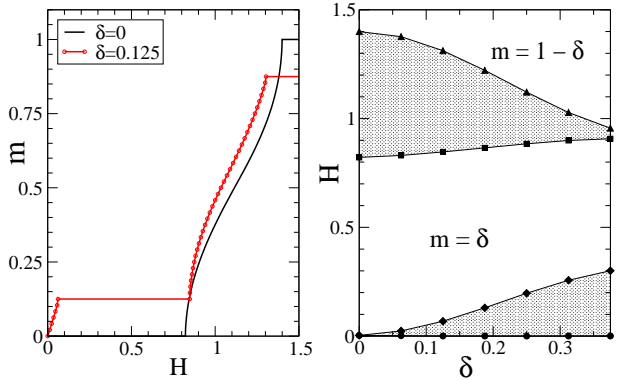}   
\caption{(Color online) Plateaus phase diagram of a 2-leg ladder for
$t_{\perp} = J_{\perp} = 1.0$ and $t_{\parallel} = J_{\parallel} =
0.2$. Doping immediately destroys the $m = 0$ plateau present at
half-filling while a large doping-dependent $m = \delta$ plateau
appears for a small critical field.}
\label{fig:large-phase}
\end{figure}

\emph{Qualitative remarks on the strong-coupling limit.} Adding
interactions in the system offers the possibility to study
magnetization plateaus due to interactions and also pairing, which
does not necessarily mean superconductivity. In this limit we have
$J_{\parallel},t_{\parallel},t_{\perp} \ll J_{\perp}$ and the pairing
energy defined as in Eq.~(\ref{eq:pairing}) is estimated to
be\cite{Troyer1996} $J_{\perp} - 2 t_{\perp} - 2 t_{\parallel}$ and
remains approximately constant under magnetic field because the
average magnetization in Eq.~(\ref{eq:pairing}) is zero. Two ground
states are possible at finite magnetization: either holes are paired
up and all the polarization is due to triplets, or holes pairs are
split apart and the partner electron on the rung holds the
polarization. The difference between the energies per rung of these
two states is:
\begin{eqnarray*}
m<\delta:\quad\Delta e &\sim& ( J_{\perp} - 2 t_{\perp} - 2
t_{\parallel}) \delta - J_{\perp} m \\ m>\delta:\quad\Delta e &\sim& (
2 J_{\perp} - 2 t_{\perp} - 2 t_{\parallel} ) \delta - 2 J_{\perp} m
\end{eqnarray*}
If $J_{\perp} / (t_{\perp}+t_{\parallel}) > 2$, a transition is then
possible from the state with paired holes to the state with unpaired
holes. The corresponding critical magnetization $ m_c = [1 - 2
(t_{\perp} + t_{\parallel} ) / J_{\perp}] \delta$ is always smaller
than $\delta$. This also gives a possible scenario for a wide $m =
\delta$ plateau due to interactions. For large $J_{\perp}$ and $m=0$,
we expect the ground state to have pairs of holes and singlets mostly
on rungs so that the system has a large spin gap, which gives a $m =
0$ plateau. However, the $m=0$ plateau is much smaller than the
half-filling spin gap because polarized spins in fact gain kinetic
energy if they are localized next to holes so that it is easier to
create them. As the magnetic field is increased, hole pairs are split
as suggested by the above argument and the partner electron of the
hole on the rung becomes polarized. When all of them are fully
polarized, we enter the $m=\delta$ plateau because the next spin
excitation one can do is to flip a singlet on a rung, which cost is
approximately $J_{\perp}$. The $m = \delta$ plateau can thus be wider
than the $m = 0$ one. This scenario corresponds to
\fig~\ref{fig:strong-coupling}, where no pairing is found in the
plateau from the computation of local densities of spins and holes
with DMRG. Such a scenario could be relevant for compounds with
lightly coupled dimers which could be doped with holes.
\begin{figure}[t]
\centering
\includegraphics[width=5cm,angle=270,clip]{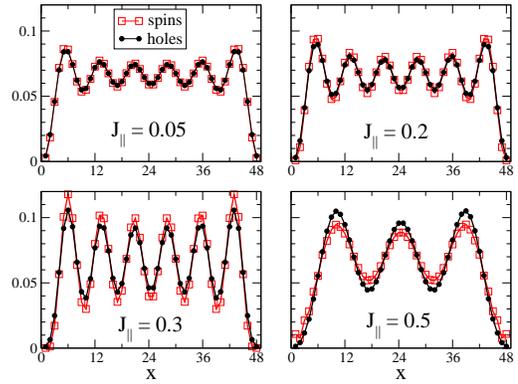}
\caption{(Color online) Local densities of holes and spins in a system
with spins in the ground state of the $m = \delta$ phase computed on a
ladder with $L = 48$ and 6 holes, $t_{\perp} = t_{\parallel} = 1.0$
and $J_{\perp} = 0.5$. As $J_{\perp}$ increases from 0 to
$J_{\parallel}$, hole pairs form in the isotropic limit. Pairing in
and above $m=\delta$ is not expected in the strong-coupling limit.}
\label{fig:transition}
\end{figure}

When one starts with unpaired holes at zero magnetic field, for
instance with $t_{\parallel}=J_{\parallel} \ll t_{\perp}=J_{\perp}$,
the large half-filled spin gap of order $J_{\perp} - J_{\parallel}$ is
immediately destroyed by doping (see \fig~\ref{fig:large-phase}).
Increasing the magnetic field further brings quickly the system into a
plateau phase with unpaired holes. While the magnetization curve is
similar to the non-interacting one below the $m=\delta$ plateau, its
width is clearly controlled by $J_{\perp}-J_{\parallel}$ rather than
by $2(t_{\perp}-2t_{\parallel})$.

An interesting feature on the pairing of holes in the $m = \delta$
plateau is the occurrence of a transition when $J_{\parallel}$ is
increased from 0 to $J_{\perp}$ from a state with unpaired holes to
paired holes (see \fig~\ref{fig:transition}). Having pairing in this
plateau is a situation that is not expected in the strong-coupling
limit as we have seen, but for intermediate $J_{\perp}$, we expect
that hopping between hole pairs and magnons can stabilize hole
pairs\cite{Roux2006} (see
Sec.~\ref{sec:band-models}). \fig~\ref{fig:transition} suggests that a
$J_{\parallel}$ comparable to $J_{\perp}$ is also needed to have
pairing in this phase.

\section{Coupled bands regime}
\label{sec:band-models}

\subsection{Bosonized Hamiltonian}
\label{sec:boson-band}

\begin{figure}[t]
\centering
\includegraphics[width=8cm]{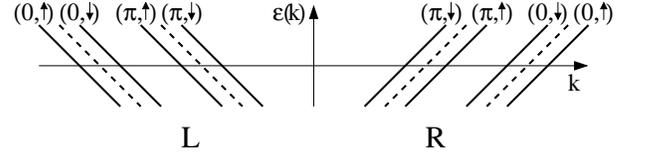}
\caption{Schematic representation of bosonized bands in the coupled
band models.}
\label{fig:schema-bos}
\end{figure}

When the interactions are not too strong compared to the interchain
hopping, it is reasonable to begin to solve the non-interacting band
structure and then add interactions. This is the approach followed in
Refs.~[\onlinecite{finkelstein_2ch, schulz_2chains, balents_2ch}].
The non-interacting band structure is simply formed of the bonding
band of energy $\varepsilon_{0, \sigma} (k)$ and an antibonding band
of energy $\varepsilon_{\pi, \sigma} (k)$. The annihilation operators
of the fermions in these bands are respectively given by:
\begin{eqnarray*}
\label{eq:new-fermions}
\psi_{0,\sigma} &=& \frac 1 {\sqrt{2}} (\psi_{1,\sigma}+\psi_{2,\sigma}) \\ 
\psi_{\pi,\sigma} &=& \frac 1 {\sqrt{2}} (\psi_{1,\sigma}-\psi_{2,\sigma})
\end{eqnarray*}
with 1 and 2 the labels of the chains. In the continuum limit, the
fermions $0$ and $\pi$ are bosonized in terms of the boson fields
$\phi_{k_y, \sigma}$ ($k_y=0,\pi$) so that we write the fermion
operator:
\begin{eqnarray}
\label{eq:bosonized-fermions}
c_{n,k_y,\sigma} & \rightarrow & \sqrt{a} \; \psi_{k_y, \sigma}(x)  \\
& = & \sqrt{a} [ e^{i k_{F,k_y}^{\sigma} x } \psi_{R,k_y,\sigma}(x) +
e^{-i k_{F,k_y}^{\sigma} x } \psi_{L,k_y,\sigma}(x) ]\nonumber
\end{eqnarray}
in which $x = n a$ ($a$ is the lattice spacing) and 
\begin{equation*}
\psi_{R/L, k_y, \sigma} = \frac{\eta_{R / L, k_y, \sigma}}{\sqrt{2 \pi
\alpha}} e^{i \epsilon_{R/L} \phi_{R/L, k_y, \sigma} }
\end{equation*} 
where $R,L$ are the labels for the right and left movers (see
\fig~\ref{fig:schema-bos}), $\alpha$ is a cutoff (typically $a$),
$\epsilon_{R/L} = \mp 1$ and the $\eta$ are Klein factors needed to
make the annihilation operators of the different fermions species
anticommute. We also have the definitions $\phi_{k_y, \sigma} = \frac
1 2 [ \phi_{L, k_y, \sigma} + \phi_{R, k_y, \sigma} ]$ while dual
fields are $\theta_{k_y, \sigma} = \frac 1 2 [ \phi_{L, k_y, \sigma} -
\phi_{R, k_y, \sigma} ] = \pi \int \Pi_{k_y, \sigma}$.

We then introduce the useful bosons $\phi_{\nu,k_y}$ with $\nu = c, s$
for charge (resp. spin) corresponding to the symmetric
(resp. antisymmetric) combination of $\phi_{ k_y, \ups}$ and $\phi_{
k_y, \downs }$. The same transformation is performed on the dual
fields, leading to the Hamiltonian:
\begin{equation*}
\label{eq:2band-free}
\Ham=\sum_{k_y=0,\pi \atop \nu=c,s} \int \frac{dx}{2\pi}\left[ u_{k_y}
(\pi \Pi_{\nu,k_y})^2 + u_{k_y} (\partial_x \phi_{\nu,k_y})^2\right],
\end{equation*} 
\noindent where $[ \phi_{\nu,k_y}(x),\Pi_{\nu',k'_y}(x')]=i
\delta_{\nu,\nu'}\delta_{k_y,k'_y} \delta(x-x')$. In the following, we
will make the usual approximation\cite{Fabrizio1993} of neglecting the
difference between the velocities $u_{k_y}$ of the $0$ and $\pi$
bands. This allows us to introduce the linear combinations:
\begin{eqnarray*}
\label{eq:band-comblin}
\phi_{c,\pm} &=&\frac 1 {\sqrt{2}} (\phi_{c,0}\pm \phi_{c,\pi}), \\
\phi_{s,\pm} &=&\frac 1 {\sqrt{2}} (\phi_{s,0}\pm \phi_{s,\pi}),
\end{eqnarray*}
with similar definitions for the dual fields $\theta_{\nu,\pm} =\pi
\int \Pi_{\nu,\pm}$. In the most general case, we have to use a $Z$
matrix to describe the evolution of the system under magnetic
field\cite{Frahm1990}. When comparing the results with the chain
models of Sec.~\ref{sec:coupled-chain}, it is useful to note that
while $\phi_{c+} = \phi_{\rho+}$ and $\phi_{s+} = \phi_{\sigma+}$ (we
use Greek letters for the fields defined in the chain models and Latin
letters for the fields defined in the band models), while there is no
simple relation between $\phi_{\rho-}$ and $\phi_{c-}$ and between
$\phi_{s-}$ and $\phi_{\sigma-}$. The magnetic field couples to the
system by a term:
\begin{equation}
\label{eq:zeeman}
\Ham = \frac{\H}{\pi} \int dx \partial_x\phi_{s+}
\end{equation}

Once interactions are turned on, two types of terms appear in the
Hamiltonian. The terms of the first type are forward scattering
interaction terms that are quadratic in the fields $\phi_{\nu,\pm}$.
The terms of the second type are backscattering interaction terms, the
expressions of which were derived in Ref.~[\onlinecite{schulz_2chains}].
Since terms containing $\cos 2\phi_{s+}$ cannot appear in the
Hamiltonian when the magnetization in nonzero, the expression of the
backscattering terms in a magnetized ladder reads:
\begin{equation}
\label{eq:backscatter-band}
\Ham_{\text{back.}}=\int dx \cos 2\theta_{c-} \left[ \frac{2g_A} {(2\pi
\alpha)^2} \cos 2\phi_{s-} + \frac{2g_B} {(2\pi \alpha)^2} \cos
2\theta_{s-}\right]
\end{equation}
From the Hamiltonian (\ref{eq:backscatter-band}), one sees that in the
ground state the field $\theta_{c-}$ is pinned to $\langle
\theta_{c-}\rangle =0$. By using the results of Ref.~[\onlinecite{orignac_suzumura}]
[Sec. 3.1 and Eqs. (20) and (56)], one can argue that in the presence
of repulsive interactions, one must have $K_{s-}<1$. Thus, one obtains
a freezing of the field related to the most relevant operator $\langle
\phi_{s-}\rangle =\frac \pi 2$.  Briefly, when $m = \delta = 0$, all
fields are massive and we have the famous spin-liquid phase, often
denoted by C0S0. When $m = 0$ but $\delta \ne 0$, the system with
repulsive interactions is in a C1S0 phase, or Luther-Emery phase (LE),
with sectors $c-$ and $s\pm$ being massive while the sector $c+$ is
massless, corresponding to the charge mode. When $m \ne 0$ and $\delta
\ne 0$, the sector $s+$ becomes massless giving rise to a C1S1
phase. Luttinger exponents associated with these sectors will be
denoted by $K_{c/s+}$. Till now, we have discussed the case of generic
nonzero magnetization.  For the specific case $m = \delta$, however, a
magnetization plateau can be expected.  Following
Ref.~[\onlinecite{Cabra2002}],
we introduce the fields
\begin{equation}
\label{eq:Cabra-connection}
\phi_{\sigma}^{\pm} = \frac 1 {\sqrt{2}} (\phi_{c, \pm} + \sigma
\phi_{s,\pm})
\end{equation}
which appear in the term that induces the opening of the $m = \delta$
plateaus
\begin{equation}
\label{eq:Cabra-term}
\int dx \cos\left[ 2 (k_{F,0}^{\sigma} + k_{F,\pi}^{\sigma} ) x -
2\sqrt{2} \phi_{\sigma}^{+}\right]
\end{equation}
with, from (\ref{eq:Luttinger}) and (\ref{eq:m}), the relation
$k_{F,0}^{\sigma} + k_{F, \pi}^{\sigma} = \pi ( n + \sigma m)$. This
term leads to the opening of plateaus when $n \pm m \in \mathbb{Z}$ as
expected in YOA theorem. This condition is a commensurability
condition which combines both spin and charge degrees of freedom. The
$m = \delta$ plateau corresponds to the locking of the
$\phi_{\ups}^{+}$ mode. The origin of the term (\ref{eq:Cabra-term})
giving rise to the plateau is the Hubbard interaction
$Un_{i\ups}n_{i\downs}$ which contains the terms
\begin{equation}
(c_{0\ups}^{\dag} c_{\pi \ups} + c_{\pi\ups}^{\dag} c_{0 \ups})
(c_{0\downs}^{\dag}c_{\pi \downs} + c_{\pi\downs}^{\dag}c_{0\downs}),
\end{equation}
which, once bosonized, yields Umklapp terms such as
\begin{equation}
\int dx \cos\left[ 2 (k_{F,0}^{\ups} + k_{F,\pi}^{\ups} ) x - 2 (
\phi_{0,\ups} +\phi_{\pi,\ups} ) \right].
\end{equation}

\subsection{Superconducting order parameters and most divergent 
fluctuations}
\label{sec:superc-band}

In this section, we define the order parameters for superconductivity
at nonzero magnetization (high field), derive their bosonized
expressions and deduce their long range correlations. Since the SU(2)
symmetry is broken by the magnetic field, we have to compute the
superconducting correlation functions $\langle \Delta_{ \sigma \sigma'
} ^{\lambda} (x) \Delta_{ \sigma \sigma' }^{\lambda \dag} (0) \rangle$
in various channels $\lambda$. We use the following microscopic
definitions for the pairing operators $\Delta_{ \sigma \sigma' }
^{\lambda} (n)$ at rung $n$:
\begin{eqnarray}
\label{eq:micro-operators} 
&\textrm{singlet: }& \Delta_{\ups \downs}^s (n) = \sum_{\sigma} \sigma
c_{r \sigma} c_{r' -\sigma}\\
\label{eq:micro-operators_Sz1} 
&\textrm{triplet: }& \left\{ \begin{array}{l} 
\Delta_{\ups \downs}^t (n) = \sum_{\sigma} c_{r \sigma} c_{r' -\sigma}\\
\Delta_{\ups \ups}^t (n) = c_{r\ups} c_{r'\ups} \\
\Delta_{\downs \downs}^t (n) = c_{r\downs} c_{r'\downs}
\end{array}\right.
\end{eqnarray}
with $r = (n,1)$ and $r' = (n,2)$ for next-nearest neighbor pairs
created on a rung and $r = (n,p)$ and $r' = (n+1,p)$ if created on the
leg $p$.  Contrarily to the case of the coupled chain regime
(Sec.~\ref{sec:coupled-chain}) where the Fermi wave-vectors are the
same in both chains, in the case of the band regime the Fermi
wave-vectors of the two bands are different ($k_{F,0}^{\sigma} \ne
k_{F,\pi}^{\sigma}$) as can be seen on \fig~\ref{fig:schema}. As a
result, a more detailed derivation of the bosonized expressions
starting from lattice expressions in a two-chain Hubbard model becomes
necessary.

\subsubsection{Bosonized form} 

Using the bosonized form (\ref{eq:bosonized-fermions}) of the fermion
operators, we can express these order parameters as a function of
products of the $\psi_{R/L, k_y, \sigma}$. In this section, only
components with the dominant contribution will be kept, i.e. we will
neglect terms of the form $\psi_R \psi_R$ and $\psi_L \psi_L$. These
dominant contributions correspond to pairing with the lowest total
momentum $q$. The case of $2k_F$ triplet pairing will be discussed in
Sec.~\ref{sec:triplet-at-2kF-band}.  At finite magnetization, the
equality of the velocities and equation (\ref{eq:m}) ensures that the
lowest momentum is $q = k^{\ups}_{F,0 / \pi} - k^{\downs}_{F,0 / \pi}
= \pi m$. To simplify the expression of the operators we will use
extensively the following results on a pinned field $\varphi$ (for
which $\langle \varphi \rangle = \textrm{cste}$): we can replace
$\langle f( \varphi ) \rangle$ by $f(\langle \varphi \rangle )$ and
the dual field has exponentially decaying correlation
functions\cite{giamarchi_book_1d}.

Starting with the interband order parameters, which read:
\begin{eqnarray*}
\label{eq:interband}
\psi_{R,0,\sigma}\psi_{L,\pi,-\sigma} &\sim& e^{i
[\theta_{c+} - \phi_{c-} -\sigma (\phi_{s+} -\theta_{s-}) ]},
\nonumber \\
\psi_{R,0,\sigma}\psi_{L,\pi,\sigma} &\sim& e^{i [\theta_{c+}
-\phi_{c-} -\sigma(\theta_{s+} -\phi_{s-}) ]},
\end{eqnarray*}
we note that they are all proportional to $e^{i \phi_{c-}}$ and thus,
since $\theta_{c-}$ is pinned, their correlations decay exponentially.
In other words, power law decay is possible only for intraband
superconducting correlations.

The intraband superconducting order parameters in the three spin
channels read respectively:\\
\noindent --Intraband singlet:
\begin{eqnarray*}
\label{eq:intraband-singlet}
\sum_\sigma \sigma \psi_{R,0,\sigma}\psi_{L,0,-\sigma} &\sim&
\sum_\sigma \sigma e^{i [\theta_{c+} + \theta_{c-} -
\sigma (\phi_{s+} + \phi_{s-})]},\\
\sum_\sigma \sigma \psi_{R,\pi,\sigma}\psi_{L,\pi,-\sigma} &\sim&
\sum_\sigma \sigma e^{i [\theta_{c+} - \theta_{c-}
- \sigma (\phi_{s+} - \phi_{s-})]}.
\end{eqnarray*}
--Intraband triplet $S^z=0$:
\begin{eqnarray*}
\label{eq:intraband-triplet-z}
\sum_\sigma \psi_{R,0,\sigma}\psi_{L,0,-\sigma} &\sim&
\sum_\sigma e^{i [\theta_{c+} + \theta_{c-} -
\sigma (\phi_{s+} + \phi_{s-})]},\\
\sum_\sigma \psi_{R,\pi,\sigma}\psi_{L,\pi,-\sigma} &\sim&
\sum_\sigma e^{i [\theta_{c+} - \theta_{c-}
- \sigma (\phi_{s+} - \phi_{s-})]}.
\end{eqnarray*}
--Intraband triplet $S^z = 1$:
\begin{eqnarray*}
\label{eq:intraband-triplet}
\psi_{R,0,\sigma}\psi_{L,0,\sigma} &\sim& e^{i
[\theta_{c+} + \theta_{c-} + \sigma (\theta_{s+} + \theta_{s-})]},\\
\psi_{R,\pi,\sigma}\psi_{L,\pi,\sigma} &\sim& e^{i
[\theta_{c+} - \theta_{c-} + \sigma (\theta_{s+} - \theta_{s-})]}.
\end{eqnarray*}

To determine which forms of superconductivity will be dominant, we
need to express the leg and rung order parameters in terms of the
intraband order parameters. We have for the rung singlet order
parameter,
\begin{equation}
\label{eq:rung-singlet-band}
\Delta_{\ups \downs}^s (x) \sim 
\sum_\sigma \frac \sigma 2 e^{i \sigma q x} ( \psi_{R,0,\sigma}
\psi_{L,0,-\sigma} - \psi_{R,\pi,\sigma} \psi_{L,\pi,-\sigma}),
\end{equation}
Whereas the rung triplet $S^z=0$ is given by:
\begin{equation}
\label{eq:rung-trip-z-band}
\Delta_{\ups \downs}^t (x) \sim 
\sum_\sigma \frac 1 2 e^{i \sigma q x} ( \psi_{R,\pi,\sigma}
\psi_{L,0,-\sigma} - \psi_{R,0,-\sigma} \psi_{L,\pi,\sigma}),
\end{equation}
therefore, since it is composed only of interband terms, 
we expect that its correlation will present exponential
decay. Lastly, for the rung triplet $S^z=1$ order parameter, we find:
\begin{equation}
\label{eq:rung-triplet-band}
\Delta_{\ups \ups}^t (x) \sim \frac 1 2 (\psi_{R,0,\ups}
\psi_{L,0,\ups} - \psi_{R,\pi,\ups} \psi_{L,\pi,\ups}),
\end{equation}
Turning to the leg singlet order parameter, we find that it reads:
\begin{equation}
\Delta_{\ups \downs}^s (x) \sim  \sum_\sigma \frac \sigma 2 e^{i
\sigma q x} \left [ e^{-i k_{F,0}^{-\sigma} a} \psi_{R,0,\sigma}
\psi_{L,0,-\sigma} + (0 \rightarrow \pi) \right ].
\end{equation}
Note that because we can neglect interband coupling, the expression is
the same on both legs. Thus leg-leg correlations on the same leg and
between the legs will have the same sign as found numerically in
\fig~\ref{fig:isocorr}. Analogously, the leg triplet $S^z=0$ operator
reads:
\begin{equation}
\Delta_{\ups \downs}^t (x) \sim \sum_\sigma \frac 1 2e^{i \sigma q x}
\left [ \sin (k_{F,0}^{\sigma} a) \psi_{R,0,\sigma}
\psi_{L,0,-\sigma} + (0 \rightarrow \pi) \right ].
\end{equation}
Note that if we take the limit $a\to 0$ this term disappears as
happened for the rung triplet. 
Finally, the leg triplet $S^z=1$ reads:
\begin{equation}
\Delta_{\ups \ups}^t (x) \sim \frac 1 2 \left [ e^{-i k_{F,0}^{\ups}
a} \psi_{R,0,\ups} \psi_{L,0,\ups} + (0 \rightarrow \pi) \right ].
\end{equation}
Since $K_{s-}<1$, we have $\langle \phi_{s-} \rangle = \frac \pi
2$. Then, all the intraband triplet $S^z=1$ have exponentially
decaying correlations. As a consequence, both the leg and the rung
triplets with $S^z=1$ have exponentially decaying correlations. The
behavior of these correlations with higher $2k_F$ momentum will be
discussed in Sec.~\ref{sec:triplet-at-2kF-band}. Another consequence
of the ordering of the field $\phi_{s-}$ is that:
\begin{eqnarray*}
\psi_{R,0,\sigma} \psi_{L,0,-\sigma} &\sim& e^{-i\frac{\pi}{2}
\sigma} e^{i (\theta_{c+} -\sigma \phi_{s+})},\\
\psi_{R,\pi,\sigma} \psi_{L,\pi,-\sigma} &\sim& e^{i\frac{\pi}{2}
\sigma} e^{i (\theta_{c+} -\sigma \phi_{s+})}.
\end{eqnarray*}
Provided that $k_{F,0}\ne k_{F,\pi}$, the leg triplet $S^z=0$ order
parameter does not vanish. We find that it has the same critical
exponent as the rung singlet order parameter, namely $\frac 1 2
(K_{c+}^{-1} + K_{s+})$. It is larger than the $m = 0$ critical
exponent $1/(2K_{c+})$ because of the appearance of the massless boson
$\theta_{s+}$. The fact that we have a finite momentum pairing with $q
= \pi m$ which is very likely to occur in a one-dimensional-like
system and is the signature of the FFLO mechanism.

\subsubsection{$S^z=1$ Triplet with a $2k_F$ momentum} 
\label{sec:triplet-at-2kF-band}

Another notable result is the presence of rung-rung triplet $S^z=1$
correlations with $2 k_F$ oscillations. Within the band
representation, it is indeed possible to find a component of the
rung-rung triplet correlations that has power law decay. Let us
consider the next terms in the expansion of the rung triplet operator:
\begin{equation*}
\psi_{1,\sigma}\psi_{2,\sigma} \sim -2 e^{i (k_{F,0}^{\sigma}+
k_{F,\pi}^{\sigma}) x} \psi_{R,0,\sigma} \psi_{R,\pi,\sigma}+\ldots
\end{equation*}
In bosonized form, this operator reads:
\begin{equation*}
\label{eq:supra-2kf}
\Delta_{\sigma \sigma,2k_F}^t(x) \sim e^{i (k_{F,0}^{\sigma}+
k_{F,\pi}^{\sigma}) x} e^{i[\theta_{c+} - \phi_{c+} + \sigma
(\theta_{s+}-\phi_{s+})]}
\end{equation*}
It has power law correlations with a critical exponent
\begin{equation*}
\frac 1 2 \left( K_{c+}+ K_{c+}^{-1} +  K_{s+}+ K_{s+}^{-1}\right)\,,
\end{equation*}
and, from equations (\ref{eq:Luttinger}) and (\ref{eq:m}), the
associated wave-vector is $ k_{F,0}^{\sigma} + k_{F,\pi}^{\sigma} = \pi
(n + \sigma m)$. Note that its critical exponent is always larger than
the one of the rung singlet order parameter.

\subsubsection{Charge-density waves order parameters}
\label{sec:band-CDW}

Here, we address the question of the CDW exponents under magnetic
field. First, the $2k_F$ CDW order parameter $n_{i}$, which vanishes
exponentially at zero magnetization, contains terms such as:
\begin{eqnarray*}
\psi_{r,0\sigma}^{\dag}\psi_{r,\pi\sigma} &\sim& e^{i
[\phi_{c-}+\sigma\phi_{s-}-r(\theta_{c-}+\sigma\theta_{s-})]}
\end{eqnarray*}
with $r=\pm$ for $R,L$ and a wave-vector $k_{F,\pi}^{\sigma} -
k_{F,0}^{\sigma}$. These order parameters decay exponentially because
they are proportional to $e^{i\phi_{c-}}$. A last possible term is
\begin{equation*}
\psi_{R,0\sigma}^{\dag}\psi_{L,\pi\sigma} \sim e^{i [\phi_{c+} -
\theta_{c-} + \sigma( \phi_{s+} - \theta_{s-})]}
\end{equation*}
with a ``$2k_F$'' wave vector $\pi(n+\sigma m)$. It decays
exponentially because of $e^{i\theta_{s-}}$. Finally, no $2k_F$ CDW
are expected under magnetic field.

Secondly, the $4k_F$ CDW order parameter\cite{White2002} is $n_i^2$
and has an exponent $2K_{c+}$ at zero magnetization. This order
parameter contains terms such as
\begin{equation*}
\psi_{R,0\sigma}^{\dag} \psi_{L,0\sigma} \psi_{R,\pi\sigma}^{\dag}
\psi_{L,\pi\sigma} \sim e^{i 2[\phi_{c+}+ \sigma \phi_{s+}]}
\end{equation*}
which have a ``$4k_F$'' wave vector $2\pi(n+m)$ and, for nonzero
magnetization, an exponent $2(K_{c+}+K_{s+})$. We also have terms like
\begin{equation*}
\psi_{R,0\sigma}^{\dag} \psi_{L,0\sigma} \psi_{R,\pi-\sigma}^{\dag}
\psi_{L,\pi-\sigma} \sim e^{i 2[\phi_{c+} + \sigma \phi_{s-}]}
\end{equation*}
with a wave-vector $2(k_{F,0}^{\sigma} + k_{F,\pi}^{-\sigma})$ and an
exponent $2K_{c+}$ which is not affected by the magnetic field. This
last term can compete with the superconducting order parameter to be
the most diverging fluctuations depending on $K_{c+}$ and $K_{s+}$.

\subsection{Interpreting the superconducting critical field $H_c$ as 
a band-filling transition}
\label{sec:band-fill-trans}

\begin{table*}[!bt]
\centering
\begin{tabular}{|c|c|c|c|}
\hline
Type of operator & fermion expression & dimension & wave-vector \\
\hline  
$S^z=1$ triplet & $\psi_{R0\uparrow}  \psi_{L0\uparrow}$
&$\frac{1}{4K_1} + \frac{1}{4K_2} +\frac{1}{2K_3}$ & 0\\ 
&  $\psi_{R\pi\uparrow}  \psi_{L\pi\uparrow}$
&$\frac{1}{4K_1} + \frac{1}{4K_2} +\frac{1}{2K_3}$ & 0\\ 
& $\psi_{R0\downarrow}  \psi_{L0\downarrow}$ 
&$\frac{1}{2K_1}+\frac{1}{2K_2}$ & 0\\
& $\psi_{R\pi\uparrow} \psi_{L0\uparrow}$ 
& $\frac{1}{4K_1} + \frac{1}{4K_2} +\frac{K_3}{2}$
& $k_{F\pi}^{\uparrow}-k_{F0}^{\uparrow}$ \\
\hline
$S^z=0$ triplet or singlet & $\psi_{R0\uparrow} \psi_{L0\downarrow}$ &
$\frac{3+\sqrt{8}}{16} (K_1^{-1}+K_2) +
\frac{3-\sqrt{8}}{16}(K_2^{-1}+K_1)+ \frac 1 8 (K_3+K_3^{-1})$ &
$k_{F0}^{\uparrow}-k_{F0}^{\downarrow}$\\ 
                          & $\psi_{R\pi\uparrow} \psi_{L0\downarrow}$ &
$\frac{3+\sqrt{8}}{16} (K_1^{-1}+K_2) +
\frac{3-\sqrt{8}}{16}(K_2^{-1}+K_1)+ \frac 1 8 (K_3+K_3^{-1})$ &
$k_{F\pi}^{\uparrow}-k_{F0}^{\downarrow}$\\ 
\hline 
\end{tabular}
\caption{The superconducting operators for the Hubbard model with an empty band
corresponding to the coupled band regime just above $\H_c$.}
\label{tab:supra}
\end{table*}

\begin{table*}[htbp]
\centering
\begin{tabular}{|c|c|c|c|}
\hline
Type of operator & fermion expression & dimension & wave-vector \\
\hline  
$SDW^z$/CDW & $\psi^\dagger_{R0\uparrow}  \psi_{L0\uparrow}$
&$\frac{K_1}{4} + \frac{K_2}{4} +\frac{K_3}{2}$ & $2k_{F0}^{\uparrow}$\\ 
 &  $\psi^\dagger_{R\pi\uparrow}  \psi_{L\pi\uparrow}$
 &$\frac{K_1}{4} + \frac{K_2}{4} +\frac{K_3}{2}$ & $2k_{F\pi}^{\uparrow}$\\ 
 & $\psi^\dagger_{R0\downarrow}  \psi_{L0\downarrow}$ 
 &$\frac{K_1}{2}+\frac{K_2}{2}$ & $2k_{F0}^{\downarrow}$\\
 & $\psi^\dagger_{R\pi\uparrow} \psi_{L0\uparrow}$ 
 & $\frac{K_1}{4} + \frac{K_2}{4} +\frac{1}{2K_3}$ 
 & $k_{F\pi}^{\uparrow}+k_{F0}^{\uparrow}$ \\
\hline
SDW$^{x,y}$ & $\psi^\dagger_{R0\uparrow} \psi_{L0\downarrow}$ &
$\frac{3-\sqrt{8}}{16} (K_1^{-1}+K_2) +
\frac{3+\sqrt{8}}{16}(K_2^{-1}+K_1)+ \frac 1 8 (K_3+K_3^{-1})$ &
$k_{F0}^{\uparrow}+k_{F0}^{\downarrow}$\\ 
& $\psi^\dagger_{R\pi\uparrow} \psi_{L0\downarrow}$ &
$\frac{3-\sqrt{8}}{16} (K_1^{-1}+K_2) +
\frac{3+\sqrt{8}}{16}(K_2^{-1}+K_1)+ \frac 1 8 (K_3+K_3^{-1})$ &
$k_{F\pi}^{\uparrow}+k_{F0}^{\downarrow}$\\ 
\hline 
\end{tabular}
\caption{The CDW/SDW operators for the Hubbard model with an empty
band corresponding to the coupled band regime just above $\H_c$.}
\label{tab:dw}
\end{table*}

In this section, we identify the observed superconducting upper
critical field $H_c$ with a band-filling transition. Such a transition
will ungap all 3 remaining sectors, leading to an enhancement of the
correlations exponents. We describe the system just above the
transition and compute the various possible exponents using
bosonization. We consider a case where the Fermi energy is positioned
in such a way that the band $(\pi,\downarrow)$ is empty while the
three other bands remain partially filled (see \fig~\ref{fig:schema}
for illustration).  The band $(\pi,\downarrow)$ being empty has
important consequences. Projecting out the high energy subspace where
the band $(0,\downarrow)$ is occupied by a single electron, one gets
in lowest order:
\begin{equation}
c^\dagger_{n,p,\downarrow} c_{n,p,\downarrow} \to \frac 1 2
c^\dagger_{n,0,\downarrow} c_{n,0,\downarrow},
\end{equation}
and as a result the on-site Hubbard interaction reduces to: 
\begin{equation}
\frac{U} 2 \sum_n n_{i,0,\downarrow} ( n_{i,0,\uparrow} +
n_{i,\pi,\uparrow}).
\end{equation}

In the absence of commensuration between the different bands, the
resulting bosonized Hamiltonian has only forward scattering
interactions. Thus, its excitations are completely gapless. The
bosonized Hamiltonian reads:
\begin{eqnarray}
\label{eq:3bands-bosonized}
\Ham&=&\Ham_0+\Ham_U \\ 
\Ham_0&=&\sum_{\nu\in \{(0,\uparrow),(0,\downarrow),(\pi,\uparrow)\}}
\int \frac{dx} {2 \pi} v_{F,\nu} \left[ (\pi \Pi_\nu)^2 + (\partial_x
\phi_\nu)^2\right] \nonumber\\
\Ham_U &=& \frac{U\alpha}{2\pi^2} \int dx \partial_x \phi_{0,\downarrow}
( \partial_x \phi_{0,\uparrow}+\partial_x \phi_{\pi,\uparrow})
\end{eqnarray}

This Hamiltonian can be fully diagonalized. In the general case, where
the Fermi velocities are all different, one needs first to perform a
rescaling: $\Pi_\nu \to \sqrt{u/v_{F,\nu}} \Pi_\nu$ and $\phi_\nu \to
\phi_\nu / \sqrt{u/v_{F,\nu}}$ where $u$ is an arbitrary quantity with
the dimension of a velocity and then diagonalize the matrix:
\begin{equation}
\left( \begin{array}{ccc} v_{F,0\uparrow}^2 & \frac{U \alpha}{4\pi^2}
\sqrt{v_{F,0\uparrow}v_{F,0\downarrow}} & 0 \\ \frac{U \alpha}{4\pi^2}
\sqrt{v_{F,0\uparrow}v_{F,0\downarrow}} & v_{F,0\downarrow}^2
& \frac{U \alpha}{4\pi^2} \sqrt{v_{F,\pi\uparrow}v_{F,0\downarrow}}\\
0 & \frac{U \alpha}{4\pi^2} \sqrt{v_{F,0\uparrow}v_{F,0\downarrow}} &
v_{F,\pi\uparrow}^2\end{array}\right)
\end{equation}

Afterwards, one performs the inverse rescaling on the diagonal matrix
to obtain the velocities of the modes and the associated Luttinger
exponents. To simplify the algebra, we will assume that all the Fermi
velocities are equal. Then, we find that the following combination of
fields
\begin{equation}
\label{eq:rotation-3band}
\left(\begin{array}{c} \phi_1 \\ \phi_2 \\ \phi_3\end{array}\right)=
\left(\begin{array}{ccc}\frac{\sqrt{2}} 2 & \frac 1 2 & \frac 1 2 \\
\frac{\sqrt{2}} 2 & -\frac 1 2 & -\frac 1 2 \\ 0 & \frac{\sqrt{2}} 2
&- \frac{\sqrt{2}} 2 \end{array} \right) \left(\begin{array}{c}
\phi_{0,\downarrow} \\ \phi_{0\uparrow} \\ \phi_{\pi \uparrow}
\end{array} \right),
\end{equation}
diagonalizes the interaction. and: 
\begin{eqnarray}
\label{eq:3band-ll-param}
u_1 K_1&=& u_2 K_2= u_3 K_3 =v_F \\ \frac {u_1} {K_1} &=& v_F
+\frac{U\alpha}{2\pi^2 \sqrt{2}}\\ \frac {u_2} {K_2} &=& v_F
-\frac{U\alpha}{2\pi^2 \sqrt{2}}\\ \frac {u_3} {K_3} &=& v_F
\end{eqnarray}
In terms of these new fields, the fermion operators read:
\begin{eqnarray*}
\label{eq:3band-fermion}
\psi_{r,0,\uparrow}&=&\frac{\eta_{r,0,\uparrow}} {\sqrt{2\pi\alpha}}
e^{i \left[ \frac 1 2 (\theta_1 - r \phi_1) - \frac 1 2 (\theta_2 - r
\phi_2)+ \frac 1 {\sqrt{2}} (\theta_3 - r \phi_3)\right]} \\
\psi_{r,\pi,\uparrow}&=&\frac{\eta_{r,\pi,\uparrow}}
{\sqrt{2\pi\alpha}} e^{i \left[ \frac 1 2 (\theta_1 - r \phi_1) -
\frac 1 2 (\theta_2 - r \phi_2)- \frac 1 {\sqrt{2}} (\theta_3 - r
\phi_3)\right]} \\ \psi_{r,0,\uparrow}&=&\frac{\eta_{r,0,\downarrow}}
{\sqrt{2\pi\alpha}} e^{i \frac{\sqrt{2}} 2\left[(\theta_1 - r \phi_1)+
(\theta_2 - r \phi_2)\right]}
\end{eqnarray*}
From these expressions, it is possible to obtain the various
superconducting order parameters and their critical exponents. The
results are gathered in table~\ref{tab:supra} and table~\ref{tab:dw}
on which one must note that the correlation exponent will be twice the
dimension of the operator.

\subsection{Numerical results on superconducting correlation functions} 
\label{sec:numerics_correlations}

\begin{figure}[t]
\centering
\includegraphics[width=7cm,angle=270,clip]{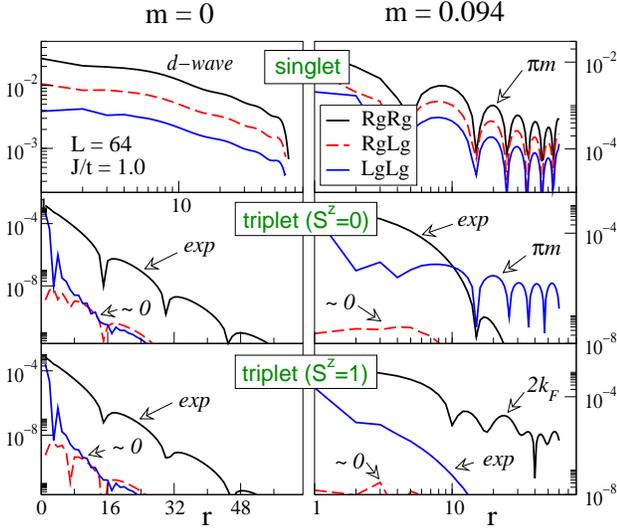}
\caption{(Color online) Absolute value of superconducting correlations
for zero and a finite magnetization in an isotropic doped ladder with
$\delta = 0.063$. Notation are ``Rg'' for rung, ``Lg'' for leg,
``$\sim 0$'' for numerically irrelevant signal, ``{\it exp}'' for
exponentially decaying correlations and ``$\pi m$'' and ``$2k_F$'' for
the wave-vectors. Note that the $S^z=1$ ``$2k_F$'' oscillations are
smoothed out by taking the absolute value. $M=1000$ states were kept
with discarded weight of order $10^{-6}$.}
\label{fig:isocorr}
\end{figure}

\begin{figure}[t]
\centering
\includegraphics[width=7cm,angle=270,clip]{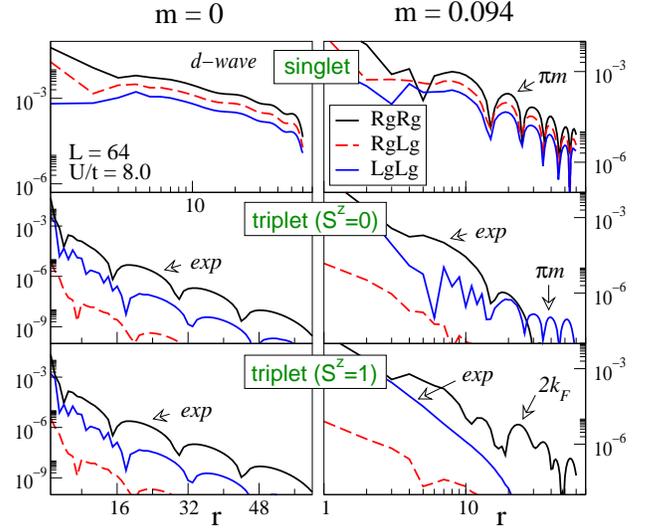}
\caption{(Color online) Same correlations as \fig~\ref{fig:isocorr}
but using the Hubbard model with $\delta = 0.063$ and $U/t =
8.0$. $M=1600$ states were kept with discarded weight of order
$10^{-6}$.}
\label{fig:isocorr_hubbard}
\end{figure}

\begin{figure}[t]
\centering
\includegraphics[width=6.5cm,angle=270,clip]{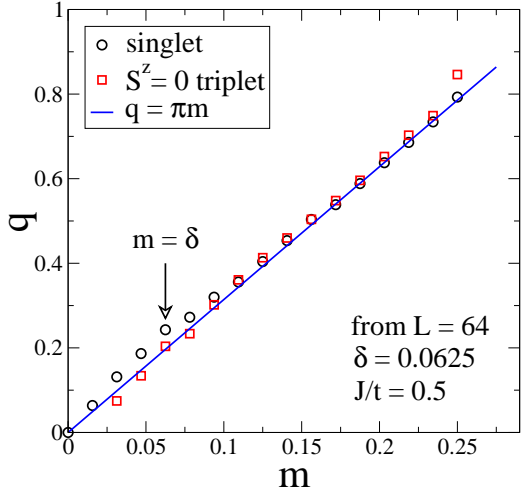}
\caption{(Color online) Verification of the $q=\pi m$ relation of
FFLO-like pairing at finite magnetization in an isotropic ladder
with $J/t=0.5$.}
\label{fig:checkq}
\end{figure}

We have computed superconducting correlation functions with DMRG using
the definitions of Eqs.~(\ref{eq:micro-operators}) and
(\ref{eq:micro-operators_Sz1}). As predicted by bosonization results,
only modified $d$-wave singlet superconductivity is found when $m = 0$
(rung-leg correlations have an opposite sign to rung-rung and leg-leg
correlations, see Ref.~[\onlinecite{White2002}]
for a bosonization discussion), while $S^z = 0$ leg-leg triplet and
$S^z = 1$ rung-rung triplet correlations emerge under high magnetic
field (see \fig~\ref{fig:isocorr} for the t-J model and
\fig~\ref{fig:isocorr_hubbard} for the repulsive Hubbard model).
Furthermore, these correlations oscillate with wave-vectors $q = \pi
m$ for the $S^z = 0$ channel and $q = 2 k_F $ for the $S^z = 1$
channel. The $q = \pi m$ relation has been checked in all the FFLO
phase (see \fig \ref{fig:phasediagram} and discussion in
Sec.~\ref{sec:phasediagram}), by fitting the oscillating correlations
with $A \cos( q r + \varphi ) / r^{\alpha}$ (see \fig
\ref{fig:checkq}), confirming the FFLO-like mechanism. We observe that
$S^z = 0$ rung-rung triplet correlations have an exponential behavior
while rung-rung singlet are algebraic, in agreement with the
prediction from bosonization. Since from the point of view of the
rotation symmetry these two order parameter transform in the same way,
the difference must be caused by their different transformation
properties under interchange of the legs. Note that because of the
$\sin(k_{F,k_y}^{\sigma} a)$ factors in the leg singlet and $S^z = 0$
leg triplet, the cancellation of the intraband terms obtained in the
case of rung order parameters is absent. This explains the observation
of power-law correlations. Concerning the orbital part of pairs,
singlet Cooper pairs have a mixed $s$-wave and $d$-wave structure and
$S^z=0$ triplet can be considered to first approximation as $p$-wave
pairs along the legs (with a symmetric superposition of the two legs).

The good agreement with bosonization predictions relies on the fact
that in the doped isotropic t-J ladder, 4 Fermi points exist with
approximatively\cite{Troyer1996} $k_{F,0} \sim 3\pi/5$ and $k_{F,\pi}
\sim 2\pi/5$ when $m = 0$ and that the assumption of equal Fermi
velocities is numerically reasonable at low hole doping. The existence
of a small leg-leg triplet component (about 100 times smaller than the
rung rung triplet) in the two leg t-J ladder considered may therefore
be explained by the persistence of a small band splitting in this
strongly coupled ladder system.

\begin{figure}[t]
\centering
\includegraphics[width=0.9\columnwidth,clip]{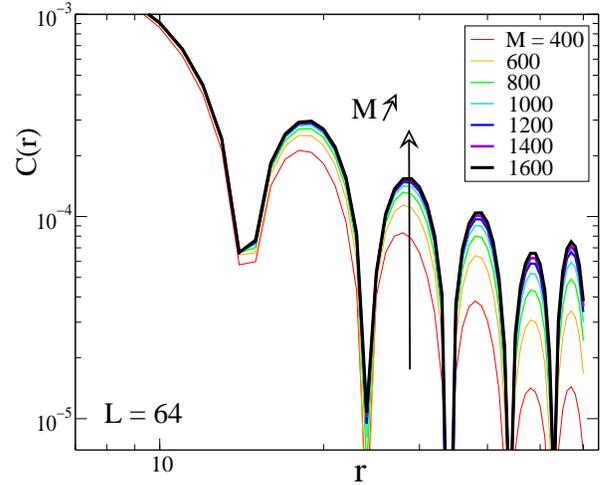}
\caption{(Color online) Convergence with the number of kept states $M$
for a fixed length $L=64$. The plotted superconducting correlation is
the singlet one and parameters are $m = 0.094$, $\delta = 0.063$ and
$J/t = 0.5$ which corresponds to the phase above the $m = \delta$
plateau and below the superconducting upper critical field $\H_c$.}
\label{fig:convM}
\end{figure}

\begin{figure}[t]
\centering
\includegraphics[width=0.9\columnwidth,clip]{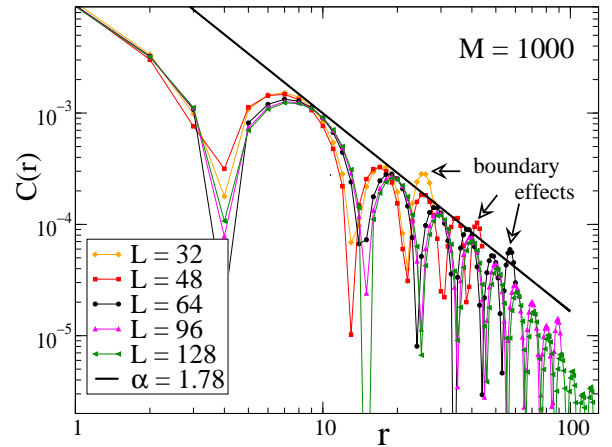}
\caption{(Color online) Convergence with size $L$ for the same
parameters as in \fig~\ref{fig:convM}, $M$ being fixed.}
\label{fig:convL}
\end{figure}

\begin{figure}[t]
\centering
\includegraphics[width=6.5cm,angle=270,clip]{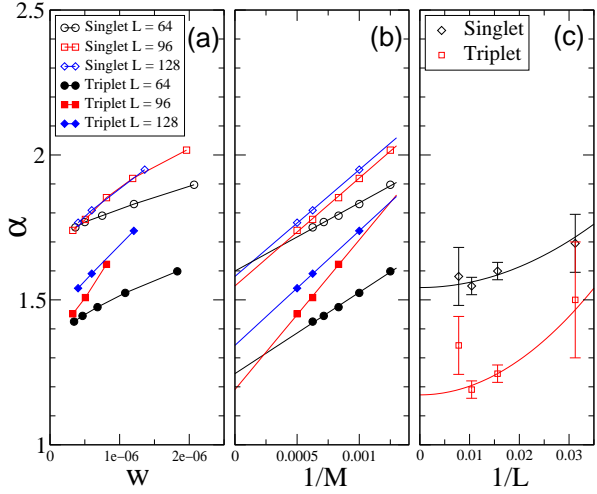}
\caption{(Color online) Extrapolation of fitted exponent as a function
of discarded weight $w$ {\bf(a)} and of the inverse of the number of
kept states $M$ {\bf(b)} for different lengths $L$. Extrapolated
results from {\bf(b)} are tentatively extrapolated vs $L$ in
{\bf(c)}. Large error bars occur when $L$ is too small (for $L=$ 32)
and because $M$ is too small (for $L =$ 128). Parameters are given in
the caption of \fig \ref{fig:convM}.}
\label{fig:convextrap}
\end{figure}

We now would like to compare precisely the algebraic decay exponents
$\alpha^s$ and $\alpha^t$ of the singlet and $S^z = 0$ leg-leg triplet
correlations. DMRG is known to often underestimate correlation
functions\cite{Schollwock2005} for a fixed number of state kept $M$.
In order to capture the behavior in the thermodynamic limit, we
computed the correlation functions for fixed $M$ ranging from 800 to
2000 and lengths $L =$ 32, 64, 96 and 128 (we worked at fixed
$\delta=1/16$ and $m=3/32$ so that these are the only accessible
cluster sizes). We extracted $\alpha(M,L)$ by fitting the data. Then,
we can extrapolate $\alpha$ in the $M \rightarrow \infty$ limit to get
a correct exponent $\alpha(L)$ at size $L$ from which we can do a
finite size scaling. On \fig~\ref{fig:convM}, we see that for a given
size, $\alpha(M)$ decreases as $M$ is enlarged, roughly like $ \alpha(
M ) \sim 1 / M$ (see \fig~\ref{fig:convextrap} {\bf(b)}) as was
previously found. We note from \fig~\ref{fig:convL} that the larger
the system, the larger $M$ is needed to reach a good
convergence. Convergence as a function of the discarded weight is also
given on \fig~\ref{fig:convextrap} and has qualitatively the same
behavior. For $L = 128$, the convergence is slower with $M$ than for
$L = 96$, probably because we would need larger $M$ to have a correct
accuracy (the $1 / M$ might be realized for large enough $M$) so that
we believe the results are not as reliable as for $L =$ 64 and 96
(larger $M$ would be very expensive numerically). On one side of \fig
\ref{fig:convextrap} {\bf(c)}, $M = 2000$ is too small (for $L = 128$)
while on the other side, it is difficult to extract $\alpha$ because
the system is too small to resolve enough oscillations (when $L =
32$). Extrapolations can be tentatively done with some uncertainties
which can be roughly estimated. Therefore, we can infer from \fig
\ref{fig:convextrap} that $\alpha^s = 1.54 \pm 0.15$ is greater than
$\alpha^t = 1.17 \pm 0.15$ which pleads in favor of dominating $S^z =
0$ triplet correlations in this part of the phase diagram (above the
$m = \delta$ plateau and below $\H_c$) but the difference is rather
small. Finite size effects could explain that the bosonization
prediction $\alpha^t = \alpha^s$ is not realized on numerical
results. Note that $\alpha < 2$ is sufficient to have a divergent
superconducting susceptibility in the system. Concerning CDW
exponents, they are difficult to compute numerically because of OBC
but we will see that we can deduce from the two-particle gap of
\fig~\ref{fig:gaps} that the system is in a superconducting phase.


\subsubsection*{Evolution of the Luttinger parameter $K_{s+}$}

\begin{figure}[t]
\centering
\includegraphics[width=0.9\columnwidth,clip]{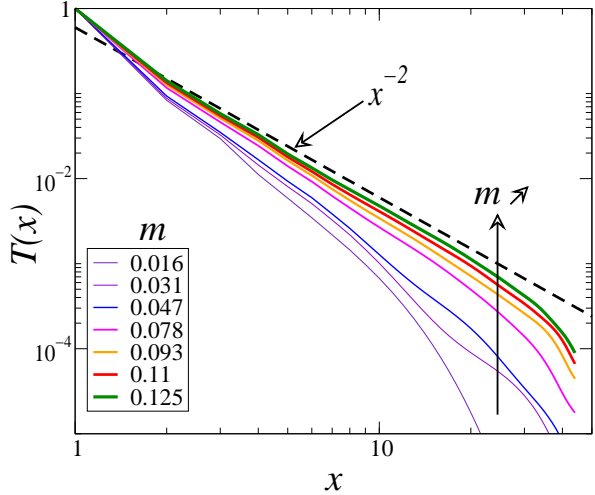}
\caption{(Color online) Normalized 4-points spin correlations
$T(x) = \moy{S^{+}_{2}(x) S^{+}_{1}(x) S^{-}_{2}(0) S^{-}_{1}(0)}$ for
various magnetization in the isotropic t-J ladder with $J/t=0.75$ and
$\delta=0.063$. Their decay exponent is $2K_{s+}^{-1}$ which gives
access to $K_{s+}(H)$. Correlations were computed on a system with $L
= 64$ and $M = 1200$ states kept.}
\label{fig:Measuring_Ks}
\end{figure}

\begin{figure}[t]
\centering
\includegraphics[angle=270,width=7cm,clip]{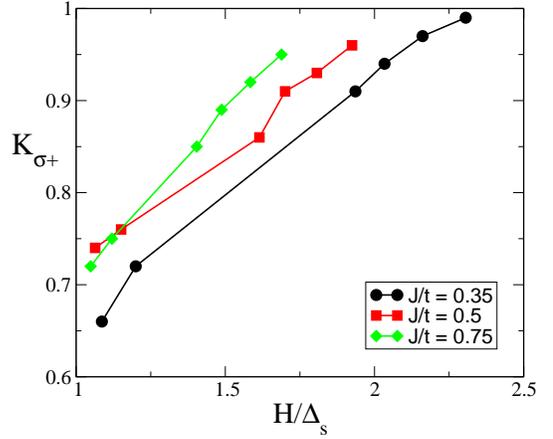}
\caption{(Color online) $K_{s+}(H / \Delta_s)$ from DMRG computations
on a system with parameters of \fig~\ref{fig:Measuring_Ks}.}
\label{fig:Ks}
\end{figure}

We now want to have access to the evolution of the Luttinger exponent
$K_{s+}$ of the gapless mode which appears at finite magnetization. We
remark that the bosonized form of the triplet creation operator on a
rung $S^{+}_{2}(x) S^{+}_{1}(x)$ contains dominant terms such as
\begin{equation}
\psi_{L,0\ups}^{\dag} \psi_{L,0\downs} \psi_{R,\pi\ups}^{\dag}
\psi_{R,\pi\downs} \sim e^{-i 2 [\phi_{s-} + \theta_{s+}]}\,.
\end{equation}
giving an exponent $2 K_{s+}^{-1}$ since $\moy{\phi_{s-}} = \pi /
2$. The wave-vector associated with this operator is $(k_{F,0}^{\ups}
- k_{F,0}^{\downs}) - (k_{F,\pi}^{\ups} - k_{F,\pi}^{\downs}) \sim 0$
since, if the difference between the Fermi velocities of the $0$ and
$\pi$ bands is negligible, we have $k^{\ups}_{F, k_y} -
k^{\downs}_{F,k_y} = \pi m$. Note that $2 K_{s+}^{-1}$ is the smaller
exponent for terms with this wave-vector. Because $K_{s+}<1$, we could
expect smaller possible exponents such as $2K_{s+}$ or $\frac 1 2 [
K_{s+} + K_{s+}^{-1}]$. For the first, one can show that fields
$2[\phi_{s+} \pm \phi_{s-}]$ or $2[\phi_{s+} +\sigma \theta_{c-}]$
cannot appear in the decomposition of $S_2^+S_1^+$. For the other,
fields $\phi_{s+} \pm \theta_{s+} \pm \theta_{c-} \pm \phi_{s-}$
cannot be decomposed in terms of the right and left movers fields with
factors $\pm$. Finally, the decay exponent of the correlations of this
order parameter is expected to be $2 K_{s+}^{-1}$ and depends only on
$K_{s+}$ and not on $K_{c+}$. Numerically, we have computed the
correlation function $\moy{ S^{+}_{2}(x) S^{+}_{1}(x) S^{-}_{2}(0)
S^{-}_{1}(0) }$. Data for the isotropic t-J model are displayed on
\fig~\ref{fig:Measuring_Ks} and show a power-law decay with an
exponent larger than $2$ and a wave-vector approaching zero. Points
are given as a function of $\H / \Delta_s$, $\Delta_s$ being the
finite-size spin gap of the system. We thus have access to the
evolution of $K_{s+}$ as a function of the magnetic field (see
\fig~\ref{fig:Ks}). The general behavior is that $K_{s+} < 1$ and
increases with magnetic field towards the limit $K_{s+} = 1$ at high
fields. Similarly, the Luttinger parameter $K_{s+}$ has been
characterized using Bethe ansatz in the context of the SO(8)
description of two-leg half-filled Hubbard ladders\cite{Konik2000}
which gave $K_{s+} < 1$. When doped, the SO(6) description\cite{Schulz1998} gives the
same constraint (results not shown) as well as an increase of $K_{s+}$ at
high magnetic field.
However, a detailed comparison of SO(6) predictions and numerics would
require to determine first the conditions and parameters at which 
both approaches match and this is beyond the scope of this article.
Lastly, the values of $K_{s+}$  obtained numerically should be slightly 
larger in the thermodynamic limit than what is found because of finite size
and finite $M$ effects as explained above.

\subsection{Generic phase diagram of the t-J model}
\label{sec:phasediagram}

In this section, we discuss the generic phase diagram in the $(\H ,
\delta)$ plane for the isotropic t-J model on the basis of DMRG and
bosonization results.

\subsubsection{Doping-dependent magnetization plateaus}
\label{sec:plateaus}

\begin{figure}[t]
\centering
\includegraphics[width=6.5cm,angle=270,clip]{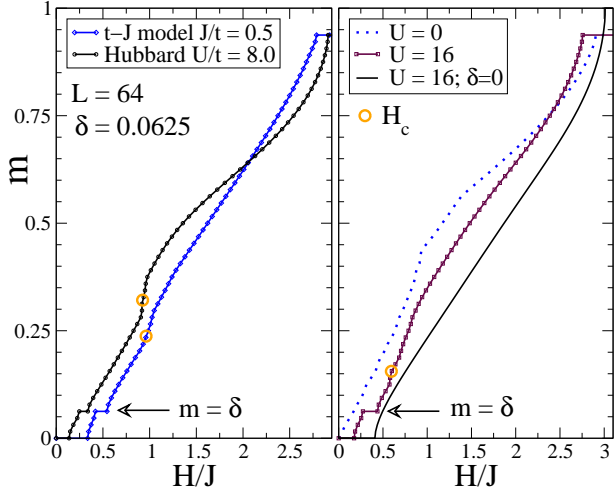}
\caption{(Color online) Magnetization curves for the isotropic t-J,
Hubbard and non-interacting systems. Irrational plateaus are clearly
visible at $m = \delta$. $\H_c$ represents the superconducting
critical field. The energy scale $J$ is defined as $4t^2/U$ in the
Hubbard model. The non-interacting system curve has been rescaled ($\H
\rightarrow \H/4$) for clarity.}
\label{fig:magnetization}
\end{figure}

\begin{figure}[t]
\centering
\includegraphics[width=6.3cm,angle=270,clip]{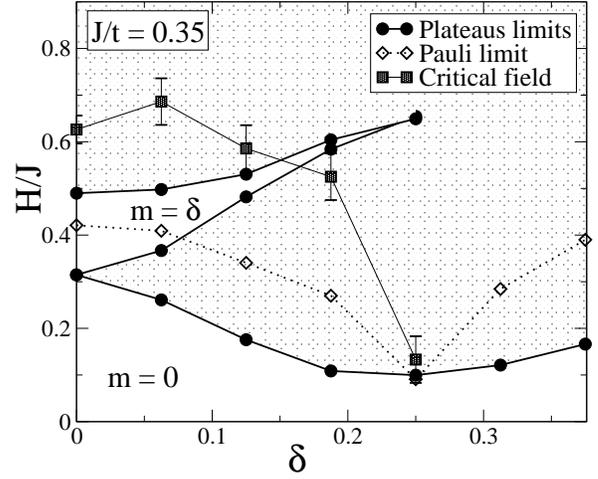}
\caption{Phase diagram for the isotropic two-leg t-J ladder in the
$(H,\delta)$ plane for $J/t = 0.35$. The large FFLO phase is delimited
by the upper limit of the $m=0$ plateau and the critical field
$\H_c$. Note also that in this part of the diagram, the $m=\delta$
phase is not expected to be a superconducting phase but rather
metallic.}
\label{fig:phasediagram}
\end{figure}

The magnetization curves $m(\H)$ of \fig~\ref{fig:magnetization},
obtained within the Hubbard and t-J models, display plateaus for $m =
0$ and $m = \delta$. Energies $E(\nh,S^z)$ were computed keeping $M =
1600$ states with the single-site method proposed by
White\cite{White2005} (we used a noise level of $10^{-6}$) at fixed
hole number $\nh$ and total magnetization $S^z$. Magnetic fields are
deduced using $\H (S^z) = E(\nh,S^z+1) - E(\nh,S^z)$, and interpolated
with $ [ \H(S^z) + \H(S^z-1) ] / 2$ if they do not belong to a
plateau. The $m = 0$ plateau simply corresponds to the well-known
spin-gap of the doped ladder. The $m = \delta$ plateau, exists at
small doping for continuous values of $\delta$ (see
\fig~\ref{fig:phasediagram} and Ref.~[\onlinecite{Roux2006}]) and thus
falls into the classes of doping-dependent irrational magnetization
plateaus predicted\cite{Cabra2000,Cabra2002} by Cabra {\it et al.}. It
can be understood as a
Commensurate-Incommensurate\cite{japaridze_cic_transition} (C-IC)
transition so that we expect the critical exponent of the
magnetization as a function of the magnetic field to be 1/2. In the
plateau phase, the mode $\phi_{\ups}^{+}$ is locked but the sector
$(\downs,+)$ remains gapless, leading to a metallic phase. Note that
$\phi_{\ups}^{+}$ is a superposition of both spin and charge modes so
that charge and spin are no more independent modes in the plateau
phase.

The Hubbard and t-J models give qualitatively the same behavior at low
magnetization. The spin gap ($m = 0$ plateau) in the Hubbard model
with $U/t = 8$ is about half of the spin gap of the t-J model with
$J/t = 0.5$ as it was previously found\cite{Jeckelmann1998}, but the
irrational plateau has roughly the same width. A larger $U$ gives a
slightly larger plateau. Both models displays a singularity of the
magnetization curve, more or less pronounced, near the superconducting
critical field $ \H_c $. The location of $\H_c$ in
\fig~\ref{fig:magnetization} have been roughly determined by looking
at local densities (see Ref.~[\onlinecite{Roux2006}] for method) since
Cooper pairs break down above this field. For larger $U$, the
superconducting critical field is smaller. These results are also
observed in the t-J model having in mind that $J/t \sim t/U$ (see
Sec.~\ref{sec:exceeding} and \fig~\ref{fig:elementary}). For higher
magnetization states, Gutzwiller projection induces a rather different
behavior between strongly interacting systems (Hubbard $U = 16$ and
t-J) the Hubbard model with $U = 8$ which displays the expected
square-root-like behaviors near critical fields in good agreement with
the band-emptying scenario proposed in
Sec.~\ref{sec:band-fill-trans}. At last, a quick comparison with the
isotropic non-interacting system proves the non-trivial role of
interactions in the apparition of these plateaus.

The magnetic part of the phase diagram of the t-J model with $J/t =
0.35$ has been computed (see \fig~\ref{fig:phasediagram}) and is very
similar to the one obtained with $J/t = 0.5$ (see
Ref.~[\onlinecite{Roux2006}]) but have larger $m = \delta$
plateaus. We propose that such a phase diagram is generic for the
isotropic t-J and Hubbard models under Zeeman effect at low doping and
for parameters $0.25 \lesssim (J/t \sim 4t/U) \lesssim 1.0$
corresponding to the strong-coupling regime. Varying $J/t$ modifies
the width of the different phases as proposed in
\fig~\ref{fig:elementary}.  Lastly, we note that $\delta = 1/4$
corresponds to the end of the plateau both for $J/t = 0.35$ and $J/t =
0.5$. Furthermore, the magnetization curves are strongly modified with
$\delta > 1/4$ (data not shown) which suggests that this hole density
corresponds to a singular point.

\subsubsection{Charge gaps and exceeding of the Pauli limit}
\label{sec:exceeding}

\begin{figure}[t]
\centering
\includegraphics[width=7cm,angle=270,clip]{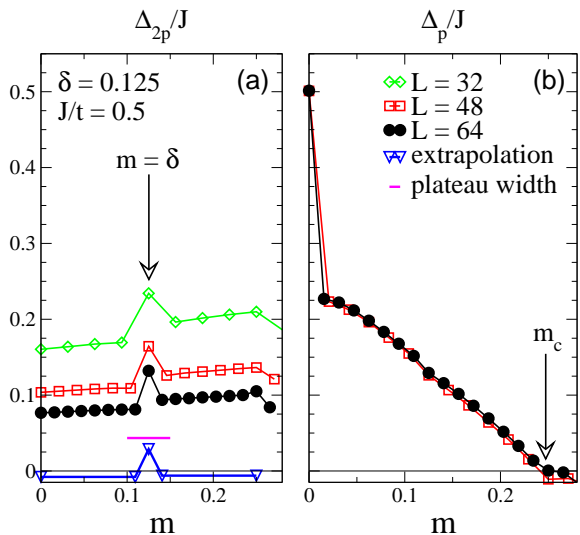}
\caption{(Color online) Two-particle gap {\bf(a)} and pairing energy
{\bf(b)} for isotropic ladders as a function of magnetization. An
anomaly in the two-particle gap at $m = \delta$ is clearly visible
while the pairing energy remains finite up to the
superconducting-metallic state transition. $m_c$ represents the
magnetization corresponding to the superconducting critical field
$H_c$.}
\label{fig:gaps}
\end{figure}

To study the charge degree of freedom of the system, we computed the
two-particle gap $\Delta_{\textrm{2p}}$ (related to the inverse
compressibility) and pairing energy $\Delta_\textrm{p}$ as a function
of the magnetization $m$ to discuss the nature of the ground
state. With $\nh$ the number of holes and $S^z$ the total spin along
the magnetic field, we have the definitions
\begin{eqnarray}
\label{eq:gaps}
\Delta_{\textrm{2p}}(S^z) &=& E(\nh+2,S^z) + E(\nh-2,S^z)\nonumber\\
 && - 2 E(\nh,S^z)\,,
\end{eqnarray}
and
\begin{eqnarray}
\label{eq:pairing}
\Delta_{\textrm{p}}(S^z) &=& E(\nh-1,S^z+1/2) + E(\nh-1,S^z-1/2)
\nonumber \\ && - E(\nh,S^z) - E(\nh - 2,S^z)\,.
\end{eqnarray}
The evolution of these gaps under increasing magnetization is
displayed on \fig \ref{fig:gaps}. Superconducting correlations are
also algebraic when the pairing energy is finite from \fig~6 of
Ref.~[\onlinecite{Roux2006}]. These observations confirm that the
system is in a superconducting state, except in the $m = \delta$
plateau where a finite two-particle charge gap is found as well as a
strong reduction of all superconducting correlations from \fig~6 of
Ref.~[\onlinecite{Roux2006}]. However, the field $\phi_{\downs}^+$
remains gapless in the irrational plateau while other fields are
gapful. This suggests that the system is in a metallic phase in this
plateau phase. Increasing the magnetic field brings the system back
into a superconducting phase, giving an example of reentrant
superconductivity which shares similarity with a proposal for a
bilayer system\cite{Buzdin2005}. The finite value of $ \Delta_{2
\textrm{p}}$ in the thermodynamic limit within the plateau phase can
be inferred by using Eq. (\ref{eq:gaps}) and the constraint $m =
\delta$ for continuous values of hole doping, which are more general
hypothesis than the particular case under study. Indeed, it is
straightforward to find that it exactly equals the width $E(\nh,S^z +
1) - E(\nh,S^z - 1)$ of the plateau in the thermodynamical limit. This
is coherent with the numerical results of \fig~\ref{fig:gaps} obtained
at fixed $\delta$. It simply means that removing a hole pair in a
system which has $m = \delta$ (i.e. the same number of hole pairs and
magnons) is energetically equivalent to adding a magnon in a system
with $m' = \delta'$ at a slightly lower density. Thus, one has to pay
the energy gap of the $m' = \delta'$ plateau for that, and this gap
equals the $m = \delta$ gap in the thermodynamical limit.

We now come to the computation of the theoretical Pauli limit $\H_p$
and of the actual superconducting critical field $\H_c$. The Pauli
limit is defined by equaling the condensation energy at $T=0$ and
$\H=0$ to the energy of the electrons coming from a broken pair and
stabilized by Zeeman effect\cite{Shimahara1997}. We have access
numerically to the condensation energy of a pair of electrons et zero
magnetization through $\Delta_{\textrm{p}} (S^z = 0)$. Paired
electrons in a singlet state do not take advantage of Zeeman effect
contrary to unpaired electrons which can be polarized along the
magnetic field. The total energy of these two electrons is
$2E(\nh-1,+1/2) - 2 \times \frac 1 2 \times \H$, to be compared with the total 
energy of the same system with paired
electrons, $E(\nh,0) + E(\nh-2,0)$. By looking at
Eq.~(\ref{eq:pairing}) with $S^z = 0$, we find that the Pauli field
simply reads $\H_p = \Delta_{\textrm{p}} (S^z = 0)$. What actually
occurs in the system is a transition to a FFLO state which changes the
nature of the ground-state so that the Pauli limit can be exceeded. The superconducting
upper critical field is deduced from the location at which
$\Delta_{\textrm{p}}(S^z)$ crosses zero which gives a critical
magnetization $m_c$ and in turn the critical field $\H_c$ reported on
\fig~\ref{fig:phasediagram}. It is important to note that these
calculations are exact and with no approximation. The
superconducting critical field $\H_c$ can also be interpreted in the
bosonization language as the emptying of the $(\pi,\downs)$ band (see
Sec. \ref{sec:band-fill-trans}). From \fig \ref{fig:phasediagram} and
\fig \ref{fig:elementary}, it is clear that Pauli limit is exceeded at
low doping and for a wide region of parameters. This exceeding can be
associated with the FFLO mechanism discussed in
Sec.~\ref{sec:numerics_correlations} from the behavior of the
superconducting correlations.

At last, we remark from \fig~\ref{fig:gaps} that the pairing energy
has a discontinuity upon magnetizing the ladder (in the $m\rightarrow
0$ limit) very similar to the known discontinuity of the spin
gap\cite{Lin1998, Poilblanc2000} upon doping the ladder (in the
$\delta \rightarrow 0$ limit). Note also that the strong reduction of
the pairing energy and of the spin gap for $\delta = 1/4$ observed on
\fig~\ref{fig:phasediagram} is related to the proximity of the CDW
phase\cite{White2002} which occurs for $J/t\sim0.2$.

\subsubsection{Phase diagram}

\begin{figure}[t]
\centering
\includegraphics[width=6.2cm,angle=270,clip]{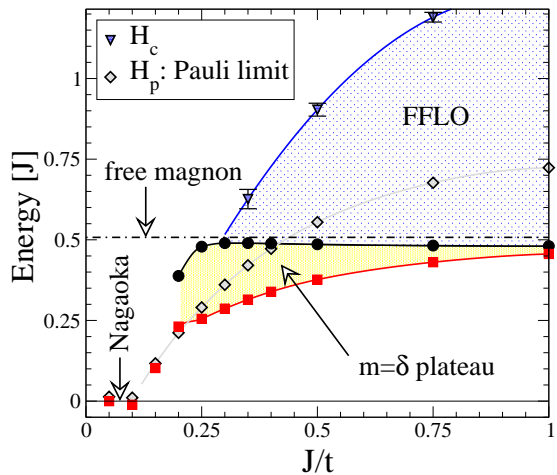}
\caption{(Color online) $\H_c$, $\H_p$ and $m=\delta$ plateau on an
isotropic two-leg t-J ladder doped with 2 holes as a function of $J/t$
(this corresponds to the $\delta \rightarrow 0$ limit of the phase
diagram of \fig~\ref{fig:phasediagram}). The red curve (squares)
corresponds to the upper limit of the $m=0$ plateau and the lower
limit of the $m = \delta$ plateau while the black curve (circles) to
the upper limit of the $m = \delta$ plateau. The FFLO phase is
delimited by the black curve (circles) and the blue curve
(triangles). Results are extrapolated from DMRG results on systems
with $L=32, 48, 64$. For small $J/t$, a Nagaoka phase is found but
because of strong finite size effects, we do not discuss this part of
the diagram. Lines are guide for the eyes.}
\label{fig:elementary}
\end{figure}

The phase diagram of \fig~\ref{fig:phasediagram} is proposed to be
generic in the coupled band regime for strong repulsive
interactions. In addition to the magnetic properties described above,
the system is in a LE superconducting state in the $m = 0$ plateau, in
a Luttinger Liquid superconducting phase below $\H_c$ and in a
metallic phase in the $m = \delta$ plateau phase. Above $\H_c$, the
system switches to a metallic phase with three ungapped sectors which
couple both spin and charge degrees of freedom. The central charge $c$
is expected to take the following values in each phase encountered as
the magnetic field is increased (taking for example a vertical cut on
\fig~\ref{fig:phasediagram} with $\delta=0.063$): $c = 1$ ($m=0$
plateau), 2 (FFLO below the $m=\delta$ plateau), 1 ($m=\delta$
plateau), 2 (FFLO above the $m=\delta$ plateau), 3 (above $\H_c$), 2
(saturation phase). Note that one could expect a last transition $c =
2 \rightarrow 1$ for certain parameters, corresponding to a situation
where the $(0,\ups)$ is completely filled. Since such a transition
would induce a cusp in the magnetization curve and there is no sign of
it on \fig~\ref{fig:magnetization}, we conclude that $c=2$ is the
value for the saturation in this regime.

To evaluate the role of $J/t$ on this generic phase diagram, we
computed various energy gaps and critical fields in the $\delta
\rightarrow 0$ limit, i.e. for two holes on a ladder with $L
\rightarrow \infty$. Results are extrapolated from systems of length
$L=$ 32, 48, 64 and are displayed on \fig~\ref{fig:elementary}. First,
we note that for $J/t$ very small, a Nagaoka phase competes the LE
phase and induces strong finite size effects in the ground states,
thus the displayed boundary between these phases \emph{does not}
correspond to the thermodynamical transition. Hence, we only focus on
points for which the LE phase is stable (on finite size systems) which
are found to be qualitatively for $0.25 \leq J/t \leq 1.0$. For large
$J/t$ the model undergoes a phase separation between holes and spins
and we are not interested in the behavior near this instability. In
this region of parameters, we have the following behavior: the width
of the plateau increases for decreasing $J/t$ (for $J/t=1.0$, the
$m=\delta$ plateau is hardly visible at finite density, data not
shown), while the exceeding of the Pauli limit increases with $J/t$
(which simply follows the increase of the pairing energy).

These elementary excitation gaps at $\delta\rightarrow 0^+$ can be
related to the dynamics of the system\cite{Poilblanc2000} when
$m=0$. The difference between $\min(\Delta_p, \Delta_M)$ and the spin
gap corresponds to the binding energy of the resonant magnetic mode.
It is interesting to note that the upper limit of the $m = \delta$
plateau is approximatively independent of $J/t$ and corresponds to a
free magnon (which is nothing but the spin gap of the undoped system
$\Delta_M \sim J/2$). Indeed, once the hole pair is bound to a magnon,
the next magnetic excitation is to create a magnon in the remaining
undoped background, which cost is $\Delta_M$, slightly renormalized by
the scattering with the holes-magnon bound-state. This supports the
phenomenological mechanism proposed in Ref.~[\onlinecite{Roux2006}]
for the opening of the irrational plateau. The binding of the hole
pair to the magnon comes from the gain in kinetic energy holes have in
a locally ferromagnetic environment. It is thus expected that the
binding energy increases with $t/J$. It is also likely to find such a
bound state in the vicinity of a Nagaoka phase which here competes the
LE phase. It also explains the decreasing of the pairing energies as
$J/t$ is reduced. Note that previous studies of this resonant magnetic
mode with exact diagonalization\cite{Poilblanc2000} and periodic
boundary conditions agree with these results, and hence suggest that
open boundary conditions do not affect this generic phase diagram.

\section{Coupled chain regime}
\label{sec:coupled-chain}

\subsection{Dominant exchange model}

By strongly reducing the interchain hopping amplitude, one can reach
the coupled chain regime in which interchain hopping is an irrelevant
perturbation\cite{wen_coupled_chains, boies_moriond,
yakovenko_manychains, brazovskii_transhop} and the relevant interchain
couplings are either the Josephson coupling (for attractive intrachain
interactions) or the exchange coupling (for repulsive interchain
interaction). An important point is that even if the bare model has
only interchain hopping, Josephson and exchange couplings are
generated by the Renormalization Group (RG) flow\cite{boies_moriond,
yakovenko_manychains, brazovskii_transhop} and, as a result, the
effective model always contain this type of interactions. In the rest
of the paper, we will call this strong coupling limit the ``chain
representation''. The bosonized Hamiltonian describing the two
uncoupled chains reads\cite{giamarchi_book_1d}:
\begin{equation}
\label{eq:2chain-free1}
\Ham=\sum_{i=1,2 \atop \nu=\rho,\sigma} \int \frac{dx}{2\pi}\left[ u_\nu
K_\nu (\pi \Pi_{\nu,i})^2 + \frac {u_\nu} {K_\nu} (\partial_x
\phi_{\nu,i})^2\right],
\end{equation}
\noindent where we have dropped terms $\propto g_{1\perp} \cos
\sqrt{8} \phi_{\sigma,i}$ since these terms are marginally irrelevant
in the case of repulsive interactions. In the case of attractive
interactions, they also become irrelevant upon the application of a
magnetic field strong enough to induce a C-IC
transition\cite{japaridze_cic_transition}.

\subsubsection{Dominant exchange}
\label{sec:dominant-exchange}

For small $J_{\parallel}/t_{\parallel}$ ratios, the analysis of the
scaling dimensions in the t-J model on a single chain\cite{ogata_tj}
shows that the dominant interchain coupling is the exchange one. This
term reads
\begin{widetext}
\begin{equation}
\label{eq:interchain-echange}
\frac{2 J_\perp \alpha}{(2\pi\alpha)^2} \cos \sqrt{2}
(\phi_{\rho,1}-\phi_{\rho,2}) \left[ \cos \sqrt{2}
(\theta_{\sigma,1}-\theta_{\sigma,2}) +\frac 1 2 \cos \sqrt{2}
(\phi_{\sigma,1} -\phi_{\sigma,2})+ \frac 1 2 \cos \sqrt{2}
(\phi_{\sigma,1} +\phi_{\sigma,2}+2\pi m x) \right],
\end{equation}
\end{widetext}
\noindent where we have the usual definition $\theta_{\nu,i}=\int \pi
\Pi_{\nu,i}$. For nonzero magnetization, the last term is oscillating
and drops from the Hamiltonian. It is convenient to introduce the new
fields
\begin{eqnarray*}
\label{eq:rotation-fields}
\phi_{\nu,\pm}&=&\frac 1 {\sqrt{2}} (\phi_{\nu,1}\pm\phi_{\nu,2}),
\\ \theta_{\nu,\pm}&=&\frac 1 {\sqrt{2}} (\theta_{\nu,1}\pm\theta_{\nu,2}).
\end{eqnarray*}
With this transformation, the exchange term is rewritten as
\begin{equation}
  \label{eq:exchange-rewritten}
  \frac{2J_\perp}{(2\pi\alpha)^2} \int dx \cos 2 \phi_{\rho-}
  \left[\cos 2\theta_{\sigma -} + \frac 1 2 \cos 2 \phi_{\sigma
  -}\right] .
\end{equation}
Moreover, the two chains being equivalent, the chain Hamiltonian is
rewritten as
\begin{equation}
\label{eq:2chain-free}
\Ham=\sum_{r=\pm \atop \nu=\rho,\sigma} \int \frac{dx}{2\pi}\left[ u_\nu
K_\nu (\pi \Pi_{\nu,r})^2 + \frac {u_\nu} {K_\nu} (\partial_x
\phi_{\nu,r})^2\right].
\end{equation}
Obviously, the Hamiltonians describing the fields $\phi_{\sigma+}$ and
$\phi_{\rho+}$ are purely quadratic indicating that the total charge
and the total spin excitations are gapless.  The fields
$\phi_{\sigma-}$ and $\phi_{\rho-}$ are described by a generalized
sine Gordon model. The sine Gordon interaction term
(\ref{eq:exchange-rewritten}) can be treated within a RG analysis. The
scaling dimensions of the term $\cos 2 \phi_{\rho-} \cos
2\theta_{\sigma -}$ is $K_{\rho-} + K_{\sigma-}^{-1}$ and scaling
dimensions of the term $\cos 2 \phi_{\rho-} \cos 2\phi_{\sigma -}$ is
$K_{\rho-} + K_{\sigma-}$. From the analysis of
Ref.~[\onlinecite{nagaosa_2ch}] we can conclude that in the case of
interest we have $K_{\sigma-}>1$ and $\cos 2 \phi_{\rho-} \cos
2\theta_{\sigma -}$ is the most relevant term and for
antiferromagnetic $J_\perp$ the ground state has $\langle
\phi_{\rho-}\rangle=0$ (mod $\pi$) and $\langle\theta_{\sigma-}\rangle
= \frac \pi 2$ (mod $\pi$).

\subsubsection{Most divergent superconducting fluctuations}
\label{sec:most-diverg-superc}

The expression of the intrachain order parameters in terms of the
fields in (\ref{eq:2chain-free}) can be found in
Ref.~[\onlinecite{giamarchi_book_1d}]. When re-expressed in terms of
the $\pm$ fields and denoting $q=\pi m$, the singlet operator reads:
\begin{eqnarray}
\label{eq:op-intrachain}
O_{SS,i}&=& \sum_{\sigma} \sigma e^{i q \sigma x}
\psi_{R,i,\sigma} \psi_{L,i,-\sigma} \\
&\sim& \sum_{\sigma} \sigma e^{i q \sigma x} e^{i(\theta_{\rho+}
-(-)^i \theta_{\rho -})} e^{-i \sigma (\phi_{\sigma +} -(-)^i
\phi_{\sigma -})},\nonumber
\end{eqnarray}
and, for triplets operators, we have:
\begin{eqnarray}
O_{TS,i,S^z=0} &=& \sum_{\sigma} e^{i q \sigma x}
\psi_{R,i,\sigma} \psi_{L,i,-\sigma} \\
&\sim& \sum_{\sigma} e^{i q \sigma x} e^{i(\theta_{\rho+} -(-)^i
\theta_{\rho -})} e^{-i \sigma (\phi_{\sigma +} -(-)^i \phi_{\sigma
-})}, \nonumber\\
O_{TS,i,S^z=1}&=& \psi_{R,i,\uparrow} \psi_{L,i,\uparrow} \\
&\sim& e^{i(\theta_{\rho+} -(-)^i \theta_{\rho -})} e^{-i
(\theta_{\sigma +} -(-)^i \theta_{\sigma -})}.\nonumber
\end{eqnarray}
For the interchain operators, the singlet operator reads: 
\begin{eqnarray}
\label{eq:op-interchain}
O'_{SS}&=& \sum_\sigma \sigma e^{i q \sigma x} \psi_{R,1,\sigma}
\psi_{L,2,-\sigma}\\ &\sim& \sum_\sigma \sigma e^{i q \sigma x}
e^{i [\theta_{\rho+} -\phi_{\rho-} + \sigma(\theta_{\sigma-}
-\phi_{\sigma+})]},\nonumber
\end{eqnarray}
while the triplet ones read:
\begin{eqnarray}
O'_{TS,S^z=0}&=& \sum_\sigma e^{i q \sigma x} \psi_{R,1,\sigma}
\psi_{L,2,-\sigma} \\ &\sim& \sum_\sigma e^{i q \sigma x} e^{i
[\theta_{\rho+} -\phi_{\rho-} + \sigma(\theta_{\sigma-}
-\phi_{\sigma+})]}, \nonumber \\
O'_{TS,S^z=1}&=& \psi_{R,1,\uparrow} \psi_{L,2,\uparrow}\\ &\sim& e^{i
(\theta_{\rho+} -\phi_{\rho-} + \theta_{\sigma+}
-\phi_{\sigma-})}.\nonumber
\end{eqnarray}
Since $\langle \phi_{\rho-}\rangle = 0$, we have $\langle
e^{i\phi_{\rho-}}\rangle \ne 0$ and by duality $\langle
e^{i\theta_{\rho-}(x)} e^{i\theta_{\rho-}(0)} \rangle \sim
e^{-x/\xi_{-}}$. This property implies that all the intrachain
superconducting correlations should decay exponentially along the
chains. On the other hand, the interchain correlations are
reinforced. The physical picture is that in this situation fermions of
opposite spin on each chain are bound together by the exchange
interaction.  Whether the dominant superconductivity is the interchain
one or the intrachain one the depends on the value of $K_{\sigma-}$.
Since we have $K_{\sigma-}>1$ in our case, the dominant
superconducting correlations are the interchain singlet and the
interchain triplet $S^z=0$. Note that these two operators have exactly
the same critical exponents.  In the case of dominant Josephson
coupling, the situation is reversed.

\subsubsection{$S^z=1$ triplet with a $2k_F$ momentum}
\label{sec:triplet-at-2kF-chain}

It is also straightforward to derive an expression of the operator
$e^{2 i k_{F,\sigma} x} \psi_{R,1,\sigma} \psi_{R,2,\sigma}$
associated to the $2k_F$ triplet correlations. This expression reads:
\begin{equation}
e^{2 i k_{F,\sigma} x} e^{i (\theta_{\rho+}-\phi_{\rho+})} e^{i \sigma
(\theta_{\sigma+}-\phi_{\sigma+})}
\end{equation}
This expression again shows power law correlations with the critical
exponent:
\begin{equation}
\frac 1 2 \left( K_{\rho+}+ K_{\rho+}^{-1} + K_{\sigma+}+
K_{\sigma+}^{-1}\right)
\end{equation}
very similar to what has been obtained in
Sec.~\ref{sec:triplet-at-2kF-band}.

\subsection{favoring the Josephson coupling in the coupled chain regime}

\begin{table*}[htbp]
\centering
\begin{tabular}{|c|c|c|}
\hline Dominant interaction & Dominant superconducting correlations &
Critical exponent \\ \hline Exchange coupling ($K_{\sigma-}>1$) & rung
singlet and rung triplet $S^z=0$ & $\frac{1}{2
K_{\rho+}}+\frac{K_{\sigma+}}{2}$ \\
Josephson coupling  ($K_{\sigma-}>1$) &   leg
triplet $S^z=1$ & $\frac{1}{2 K_{\rho+}}+\frac{1}{2K_{\sigma+}}$ \\
\hline 
\end{tabular}
\caption{The different dominant superconducting fluctuations with the
associated critical exponents. Josephson coupling dominates for
$K_{\rho-}>1$ and exchange coupling dominates for $K_{\rho-}<1$.}
\label{tab:coupled-chains}  
\end{table*}

Let us consider the regime of small interchain hopping and assume that
now the intra-chain terms are such that $J_{\parallel} \sim 4
t_{\parallel}$. In this case, the dominant term becomes the Josephson
coupling\cite{ogata_tj} and the perturbation term
Eq.~(\ref{eq:exchange-rewritten}) is replaced by:
\begin{widetext}
\begin{equation}
\label{eq:josephson}
\frac{2\lambda \alpha}{(2\pi\alpha)^2} \cos \sqrt{2}
(\theta_{\rho,1}-\theta_{\rho,2}) \left[ \cos \sqrt{2}
(\theta_{\sigma,1}-\theta_{\sigma,2}) +\frac 1 2 \cos \sqrt{2}
(\phi_{\sigma,1} -\phi_{\sigma,2})+ \frac 1 2 \cos \sqrt{2}
(\phi_{\sigma,1} +\phi_{\sigma,2}+2\pi m x) \right].
\end{equation}
\end{widetext}
The treatment parallels the one of the exchange coupling in
Sec.~\ref{sec:dominant-exchange}. The Josephson term is rewritten in
the form:
\begin{equation}
\label{eq:josephson-rewritten}
\frac{2\lambda}{(2\pi\alpha)^2} \int dx \cos 2 \theta_{\rho-}
\left[\cos 2\theta_{\sigma -} + \frac 1 2 \cos 2 \phi_{\sigma
-}\right].
\end{equation}
The scaling dimensions of the term $\cos 2 \theta_{\rho-} \cos
2\theta_{\sigma -}$ is $K_{\rho-}^{-1} + K_{\sigma-}^{-1}$ and scaling
dimensions of the term $\cos 2 \phi_{\rho-} \cos 2\phi_{\sigma -}$ is
$K_{\rho-}^{-1} + K_{\sigma-}$.  A similar RG analysis to the one of
Sec.~\ref{sec:dominant-exchange} yields for $K_{\sigma-} >1$ $\langle
\theta_{\rho-}\rangle=0$ (mod $\pi$) and
$\langle\theta_{\sigma-}\rangle=\frac \pi 2$ (mod $\pi$) and for
$K_{\sigma-} >1$ $\langle \theta_{\rho-}\rangle=0$ (mod $\pi$) and
$\langle \phi_{\sigma-}\rangle=\frac \pi 2$ (mod $\pi$).  This time,
interchain correlations are decaying exponentially, and the dominant
superconducting correlations are the intrachain ones. Since
$K_{\sigma-} >1$, $\langle \theta_{\sigma-}\rangle =\frac \pi 2$ and
the intrachain $S^z=1$ triplet superconductivity is dominant.  For
$K_{\sigma-}<1$ the dominant superconducting correlations are the
intrachain singlet and intrachain triplet $S^z=0$.

The results are summarized in the table~\ref{tab:coupled-chains} where
the critical exponents $\eta_\alpha$ are defined by $\langle
O_\alpha(x) O_\alpha(0)\rangle \sim |x|^{-\eta_\alpha}$. We note that
triplet $S^z=1$ superconducting correlations can never coexist with
singlet superconducting correlations.

We have tried to observe these predictions numerically but using both
the t-J model with small $t_{\perp},J_{\perp}$ and Hubbard model with
small $t_{\perp}$. Unfortunately, for large $J_{\parallel} /
t_{\parallel}$, the system with open boundary conditions has edge
effects due to the proximity of the phase separation which occurs
generically in the t-J model for large $J/t$. When edge effects are
absent and in the Hubbard model, we did not find evidences of the
proposed predictions, i.e. we mostly found cases related to the
coupled bands regime fixed point. Still, contrary to the isotropic
case which has been widely studied numerically at zero magnetization,
a systematic study of the phase diagram at zero magnetization would be
necessary before tackling the system under magnetic field. This
systematic study is out of the scope of the present article.

\section{Conclusion}

In conclusion, we studied the case of two fermionic coupled chains, or
ladders, under a magnetic field inducing a Zeeman effect in the
system. The first situation we addressed was the free and
strong-coupling limits. We found that large doping-dependent
magnetization plateaus occur for ``coupled-dimers'' systems (with
large interactions on the rungs) and that pairing is not expected in
the $m=\delta$ magnetization plateau phase. For a system with
isotropic couplings, we showed that $m=\delta$ plateaus also exist and
but pairing survives to much higher magnetizations. Furthermore, the
computation of the one and two-particles gap and bosonization
interpretation proved that the plateau phase is metallic while the
system is in a superconducting state below and above this plateau. In
addition, we computed the superconducting upper critical field which
is much larger than the Pauli limit for a wide range of the parameters
of the t-J model. Superconducting correlation functions precisely
agree with the bosonization predictions up to the accuracy of our
methods. Turning to the coupled-chain regime, other interesting
unconventional behaviors are predicted for the superconducting
correlations with, for instance, the possibility of having polarized
triplet correlations under high magnetic fields. However, quick
numerical investigations study were not able to find good parameters
providing evidences of such predictions.

\emph{Discussion on experiments} -- We now briefly discuss
consequences for experiments. First of all, the possibility of
measuring irrational magnetization plateau would give a direct access
to the hole doping $\delta$ which has recently been estimated\cite{Piskunov2005}
to be $\delta \sim 0.1$ in the superconducting phase of
doped ladders. However, the magnetization is a global
measurement including the contribution of all subsystems. For
instance, the magnetization curve of SCCO has very recently been
measured\cite{Klingeler} showing that the main contribution comes from the
chains subsystem. On the other hand, measurements such as NMR
under high magnetic field can provide local information on each subsystem.

The superconducting critical field $\H_c(T)$ of SCCO has been measured
under high pressure\cite{Braithwaite2000, Nakanishi2005} and displays
a strong anisotropy and an anomalous $\H_c(T)$ curvature. Furthermore, $\H_c(T)$ much larger than
standard Pauli limit are found, suggesting an exceeding of the
latest. The nature of the superconductivity in SCCO is a long-standing
debate because of its complex structure and the requirement of high
pressure experiments. Questions such as the nature of the pairing or
the dimensionality of superconductivity have not reached full
agreement yet. One possibility is that superconducting fluctuations of
the RVB type develops in the ladder subsystem leading to true
superconducting order once these ladders are coupled through Josephson
coupling ($T_c$ being controlled by this coupling rather than by the
magnetic energy scale $J$). In this case our study suggests that the
FFLO mechanism is relevant at the ladder subsystem level and thus could be
stabilized when coupling the ladders. The other possibility would be
that the superconducting phase is really
two-dimensional\cite{Nagata1998, Piskunov2004}, allowing other
explanations of the exceeding of Pauli limit such as triplet 
pairing\cite{Nakanishi2005} \emph{when \H=0}. NMR 
measurements\cite{Fujiwara2003} also showed that
superconductivity survives under rather high field and $p-$wave
superconductivity was suggested. Note that our study also proposes the
possibility for the emergence at high fields of pairing channels not
present at $\H = 0$ since magnetic fluctuations from which pairing
originates are strongly affected by the magnetic field.  Lastly, we
mention that anomalous curvature of $\H_c(T)$ as found in experiments
were predicted in the mean-field approach of FFLO states in quasi-one
and two-dimensional $d$-wave superconductors\cite{Shimahara1997}.

\acknowledgements

G.R. would like to thank IDRIS (Orsay, France) and CALMIP (Toulouse,
France) for use of supercomputer facilities. G.~R., P.~P. and
D.~P. thank Agence Nationale de la Recherche (France) for support.


\begin{thebibliography}{23}

\bibitem{FFLO} P.~Fulde and R.~A. Ferrell, Phys. Rev. \textbf{135},
A550 (1964); A.~I. Larkin and Y.~N. Ovchinnikov, Sov. Phys. JETP
\textbf{20}, 762 (1965); R.~Casalbuoni and G.~Nardulli,
Rev. Mod. Phys. \textbf{76}, 263 (2004).

\bibitem{Ishiguro2002} T.~Ishiguro in \emph{High Magnetic Fields},
 vol. 595 of \emph{Lecture notes in physics (Springer, 2002)}.

\bibitem{Dupuis1995}
N.~Dupuis, Phys. Rev. B \textbf{51}, 9074 (1995).

\bibitem{Shimahara1997}
H.~Shimahara and D.~Rainer, J. Phys. Soc. Jpn. \textbf{66}, 3591
(1997).

\bibitem{Sigrist2005} M.~Sigrist, AIP Conf. Proc. \textbf{789}, 165
 (2005).

\bibitem{LadderReview} E.~Dagotto and T.~M. Rice, Science \textbf{271},
618 (1996); T.~M. Rice, Z. Phys. B \textbf{103}, 165 (1997);
E.~Dagotto, Rep. Prog. Phys. \textbf{62}, 1525 (1999).

\bibitem{Cabra1997} D.~C. Cabra, A.~Honecker, and P.~Pujol,
 Phys. Rev. Lett. \textbf{79}, 5126 (1997); Phys. Rev. B \textbf{58},
 6241 (1998).

\bibitem{Chaboussant} G.~Chaboussant, M.-H.~Julien, Y.~Fagot-Revurat,
L.~P. L{\'e}vy, C.~Berthier, M.~Horvati{\'c}, and O.~Piovesana,
Phys. Rev.  Lett. \textbf{79}, 925 (1997); G. Chaboussant,
Y.~Fagot-Revurat, M.-H.~Julien, M.~E. Hanson, C.~Berthier,
M.~Horvati{\'c}, L.~P. L{\'e}vy, and O.~Piovesana,
Phys. Rev. Lett. \textbf{80}, 2713 (1998).

\bibitem{Watson2001} B.~C. Watson, V.~N. Kotov, M.~W. Meisel,
D.~W. Hall, G.~E. Granroth, W.~T. Montfrooij, S.~E. Nagler,
D.~A. Jensen, R.~Backov, M.~A. Petruska, G.~E. Fanucci, and
D.~R. Talham, Phys. Rev. Lett. \textbf{86}, 5168 (2001).

\bibitem{Landee2001} C.~P. Landee, M.~M. Turnbull, C.~Galeriu,
J.~Giantsidis, and F.~M. Woodward, Phys. Rev. B \textbf{63}, 100402(R)
(2001).

\bibitem{Frahm1999} H.~Frahm and C.~Sobiella,
 Phys. Rev. Lett. \textbf{83}, 5579 (1999).

\bibitem{Cabra2000} D.~C. Cabra, A.~De~Martino, A.~Honecker, P.~Pujol,
 and P.~Simon, Phys. Lett. A \textbf{268}, 418 (2000).

\bibitem{Cabra2001} D.~C. Cabra, A.~De~Martino, A.~Honecker, P.~Pujol,
 and P.~Simon, Phys. Rev. B \textbf{63}, 094406 (2001).

\bibitem{Hayward1995} C.~A. Hayward, D.~Poilblanc, R.~M. Noack,
D.~J. Scalapino, and W.~Hanke, Phys. Rev. Lett. \textbf{75}, 926
(1995).

\bibitem{Poilblanc2003} E. Orignac and D. Poilblanc, Phys. Rev. B
{\bf68}, 052504 (2003); D. Poilblanc, D.~J. Scalapino and S. Capponi,
Phys. Rev. Lett. {\bf91}, 137203 (2003); D. Poilblanc and
D.~J. Scalapino, Phys. Rev. B {\bf71}, 174403 (2005);

\bibitem{White2002} S.~R. White, I. Affleck, and D.~J. Scalapino,
Phys. Rev. B \textbf{65}, 165122 (2002).

\bibitem{Anderson1987} P.~W. Anderson, Science \textbf{235}, 1196
 (1987).

\bibitem{Gozar2005} A.~Gozar and G.~Blumberg, \emph{Frontiers in
 Magnetic Materials (Spinger-Verlag, 2005)}, pp. 653-695.

\bibitem{Uehara1996} M.~Uehara, T.~Nagata, J.~Akimitsu, H.~Takahashi,
N.~M\^ori, and K.~Kinoshita, J. Phys. Soc. Jpn. \textbf{65}, 2764
(1996); D.~Jerome, P. Auban-Senzier, and Y.~Piskunov in
\emph{High Magnetic Fields}, vol. 595 of \emph{Lecture notes in
physics (Springer, 2002)}.

\bibitem{Braithwaite2000} D.~Braithwaite, T.~Nagata, I.~Sheikin,
H.~Fujino, J.~Akimitsu, and J.~Flouquet, Solid State
Com. \textbf{114}, 533 (2000).

\bibitem{Nakanishi2005} T.~Nakanishi, N.~Motoyama, H.~Mitamura,
N.~Takeshita, H.~Takahashi, H.~Eisaki, S.~Uchida, and N.~Mori,
Phys. Rev. B \textbf{72}, 054520 (2005).

\bibitem{Matsukawa2004} M.~Matsukawa, Y.~Yamada, M.~Chiba,
H.~Ogasawara, T.~Shibata, A.~Matsushita, and Y.~Takano, Physica C
\textbf{411}, 101 (2004).

\bibitem{Roux2006} G.~Roux, S.~R. White, S.~Capponi, and D.~Poilblanc,
 Phys. Rev. Lett. \textbf{97}, 087207 (2006).

\bibitem{White1992} S.~R. White, Phys. Rev. Lett. \textbf{69}, 2863
  (1992); Phys. Rev. B \textbf{48}, 10345 (1993).

\bibitem{Schollwock2005} U.~Schollw\"ock, Rev. Mod. Phys. \textbf{77},
 259 (2005).

\bibitem{Cabra2002} D.~C. Cabra, A.~De~Martino, P.~Pujol, and
 P.~Simon, Europhys. Lett. \textbf{57}, 402 (2002).

\bibitem{Pruschke1992} T.~Pruschke and H.~Shiba, Phys. Rev. B
 \textbf{46}, 356 (1992).

\bibitem{Yang2001} K.~Yang, Phys. Rev. B \textbf{63}, 140511(R) (2001).

\bibitem{Yamanaka1997} M.~Yamanaka, M.~Oshikawa, and I.~Affleck,
 Phys. Rev. Lett. \textbf{79}, 1110 (1997); M.~Oshikawa, M.~Yamanaka,
 and I.~Affleck, Phys. Rev. Lett. \textbf{78}, 1984 (1997);
 P.~Gagliardini, S.~Haas, and T.~M. Rice, Phys. Rev. B \textbf{58},
 9603 (1998).

\bibitem{nagaosa_2ch} N.~Nagaosa, Sol. State Comm. \textbf{94}, 495
  (1995); N.~Nagaosa and M.~Oshikawa, J. Phys. Soc. Jpn. \textbf{65},
  2241 (1996).

\bibitem{khveshenko_2chain} D.~V. Khveshchenko and T.~M. Rice,
 Phys. Rev. B \textbf{50}, 252 (1994).

\bibitem{finkelstein_2ch} A.~M. Finkelstein and A.~I. Larkin,
 Phys. Rev. B \textbf{47}, 10461 (1993).

\bibitem{schulz_2chains} H.~J. Schulz, Phys. Rev. B \textbf{53}, R2959
 (1996).

\bibitem{balents_2ch} L.~Balents and M.~P.~A. Fisher, Phys. Rev. B
 \textbf{53}, 12133 (1996).

\bibitem{Troyer1996} M.~Troyer, H.~Tsunetsugu, and T.~M. Rice,
 Phys. Rev. B \textbf{53}, 251 (1996).

\bibitem{Fabrizio1993} M.~Fabrizio, Phys. Rev. B \textbf{48}, 15838
 (1993).

\bibitem{Frahm1990} H.~Frahm and V.~E. Korepin, Phys. Rev. B
 \textbf{42}, 10553 (1990); Phys. Rev. B \textbf{43}, 5653 (1991).

\bibitem{orignac_suzumura} E.~Orignac and Y.~Suzumura, Eur. Phys. J. B
 \textbf{23}, 57 (2001).

\bibitem{giamarchi_book_1d} T.~Giamarchi, \emph{Quantum Physics in one
 Dimension}, vol. 121 of \emph{International series of monographs on
 physics (Oxford University Press, Oxford, UK, 2004)}.

\bibitem{Konik2000}
R.~Konik, F.~Lesage, A.~W.~W. Ludwig and H. Saleur,
Phys. Rev. B \textbf{61}, 4983 (2000).

\bibitem{Schulz1998} H.~Schulz (1998), cond-mat/9808167.

\bibitem{White2005} S.~R. White, Phys. Rev. B \textbf{72}, 180403(R)
 (2005).

\bibitem{japaridze_cic_transition} G.~I. Dzhaparidze and
 A.~A. Nersesyan, JETP Lett. \textbf{27}, 334 (1978); V.~L. Pokrovsky
 and A.~L. Talapov, Phys. Rev. Lett. \textbf{42}, 65 (1979);
 H.~J. Schulz, Phys. Rev. B \textbf{22}, 5274 (1980).

\bibitem{Jeckelmann1998} E.~Jeckelmann, D.~J.~Scalapino, and
 S.~R.~White, Phys. Rev. B \textbf{58}, 9492 (1998).

\bibitem{Buzdin2005} A.~Buzdin, S.~Tollis, and J.~Cayssol,
 Phys. Rev. Lett. \textbf{95}, 167003 (2005).

\bibitem{Lin1998} H.~H. Lin, L.~Balents, and M.~P.~A. Fisher,
 Phys. Rev. B \textbf{58}, 1794 (1998).

\bibitem{Poilblanc2000} D.~Poilblanc, O.~Chiappa, J.~Riera, S.~R.~White,
 and D.~J.~Scalapino, Phys. Rev. B \textbf{62}, 14633 (2000);
 D.~Poilblanc, E.~Orignac, S.~R.~White, and S.~Capponi, Phys. Rev. B
 \textbf{69}, 220406(R) (2004).

\bibitem{wen_coupled_chains} X.~G. Wen, Phys. Rev. B \textbf{42}, 6623
 (1990).

\bibitem{boies_moriond} D.~Boies, C.~Bourbonnais, and
A.-M.~S. Tremblay, in \emph{Proceedings of the XXXIst Rencontres de
Moriond, edited by T.~Martin, G.~Montambaux, and J.~Tran Thanh Van
(\'Editions Fronti\`eres, Gif sur Yvette, France, 1996)}.

\bibitem{yakovenko_manychains} V.~M. Yakovenko, JETP
 Lett. \textbf{56}, 510 (1992).

\bibitem{brazovskii_transhop} S.~Brazovskii and V.~Yakovenko, J. de
 Phys. (Paris) Lett. \textbf{46}, L111 (1985).

\bibitem{ogata_tj} M.~Ogata, M.~U. Luchini, S.~Sorella, and
 F.~F. Assaad, Phys. Rev. Lett. \textbf{66}, 2388 (1991).

\bibitem{Piskunov2005} 
Y. Piskunov, D.~J{\'e}rome, P.~Auban-Senzier, P.~Wzietek, and A.~Yakubovsky,
 Phys. Rev. B \textbf{72}, 064512 (2005).

\bibitem{Klingeler} R.~Klingeler {\it et al.}, Phys. Rev. B
\textbf{72}, 184406 (2005); Phys. Rev. B \textbf{73}, 014426 (2006).

\bibitem{Nagata1998} T.~Nagata, M.~Uehara, J.~Goto, J.~Akimitsu,
 N.~Motoyama, H.~Eisaki, S.~Uchida, H.~Takahashi, T.~Nakanishi, and
 N.~M\^ori, Phys.  Rev. Lett. \textbf{81}, 1090 (1998).

\bibitem{Piskunov2004}
Y.~Piskunov, D.~J{\'e}rome, P.~Auban-Senzier, P.~Wzietek, and A.~Yakubovsky, 
Phys. Rev. B \textbf{69}, 014510 (2004); N.~Motoyama,
H.~Eisaki, S.~Uchida, N.~Takeshita, N.~M\^ori, T.~Nakanishi and H.~Takahashi,
Europhys. Lett. \textbf{58}, 758 (2002).

\bibitem{Fujiwara2003} N.~Fujiwara, N.~M\^ori, Y.~Uwatoko,
 T.~Matsumoto, N.~Motoyama, and S.~Uchida,
 Phys. Rev. Lett. \textbf{90}, 137001 (2003); J. of Phys.:
 Condens. Matter \textbf{17}, S929 (2005).

\end{thebibliography}
\end{document}